\documentclass[twocolumn]{aa} 

\usepackage{ltxtable}
\usepackage{longtable}
\usepackage{supertabular}
\usepackage{graphicx}
\usepackage{rotating}
\usepackage{pdflscape}
\usepackage{color}
\usepackage{natbib}
\usepackage{amssymb}
\usepackage{lscape}
\usepackage{txfonts}
\bibpunct{(}{)}{;}{a}{}{,} 

\def\teff{$T_{\mathrm{eff}}$}

\def\kms{km~s$^{-1}$}

\def\A{\AA}

\begin{document}

\title{The Gaia-ESO Survey: The inner disc, intermediate-age open cluster Pismis 18
  \thanks{Table 2 is only available at the CDS via anonymous ftp to cdsarc.u-strasbg.fr (130.79.128.5) or via http://cdsarc.u-strasbg.fr/viz-bin/qcat?J/A+A/vol/page }}

\titlerunning{The Gaia-ESO Survey: open cluster Pismis 18}

\author{D. Hatzidimitriou\inst{1,2}, E.~V. Held\inst{3}, E. Tognelli\inst{4,5}, A. Bragaglia\inst{6}, L. Magrini\inst{7}, L. Bravi\inst{7}, K. Gazeas\inst{1}, A. Dapergolas\inst{2}, A. Drazdauskas\inst{8}, E. Delgado-Mena\inst{9}, E.D. Friel\inst{10}, R. Minkevi\v{c}i\={u}t\.{e}\inst{8}, R. Sordo\inst{3}, G. Tautvai\v{s}ien\.{e}\inst{8}, G. Gilmore\inst{11}, S. Randich\inst{7}, S. Feltzing\inst{12}, A. Vallenari\inst{3}, E.~J. Alfaro\inst{13}, E. Flaccomio\inst{14}, A.~C. Lanzafame\inst{15}, E. Pancino\inst{6}, R. Smiljanic\inst{16}, A. Bayo\inst{17}, M. Bergemann\inst{18}, G. Carraro\inst{19}, A.~R. Casey\inst{20,21}, M.~T. Costado\inst{22}, F. Damiani\inst{14}, E. Franciosini\inst{7}, A. Gonneau\inst{11},  P. Jofré\inst{23},  J. Lewis\inst{11}, L. Monaco\inst{24}, L. Morbidelli\inst{6}, C.~C. Worley\inst{11}, S. Zaggia\inst{3}}

\institute{Section of Astrophysics, Astronomy and Mechanics, Department of Physics, National and Kapodistrian University of Athens, 15784 Athens, Greece
  (e-mail: dh@physics.uoc.gr)
\and
IAASARS, National Observatory of Athens, 15236 Penteli, Greece
\and
INAF, Osservatorio Astronomico di Padova, Vicolo dell'Osservatorio 5, 35122 Padova, Italy
\and 
Dipartimento di Fisica ‘E.Fermi’, Universit\'a di Pisa, Largo Bruno Pontecorvo 3, I-56127 Pisa, Italy
\and
INFN, Sezione di Pisa, Largo Bruno Pontecorvo 3, I-56127 Pisa, Italy
\and 
INAF, Osservatorio di Astrofisica e Fisica dello Spazio di Bologna, Via Gobetti 93/3, 40129, Bologna, Italy
\and
INAF, Osservatorio Astrofisico di Arcetri, Largo E. Fermi 5, 50125 Firenze, Italy
\and
Institute of Theoretical Physics and Astronomy, Vilnius University, Saul\.{e}tekio av. 3, 10257 Vilnius, Lithuania
\and
Instituto de Astrof\'isica e Ci\^encias do Espa\c{c}o, Universidade do Porto, CAUP, Rua das Estrelas, PT4150-762 Porto, Portugal
\and
Astronomy Department, Indiana University Bloomington, Swain West 319, 727 East 3rd Street, Bloomington, IN 47405-7105, USA
\and
Institute of Astronomy, Cambridge University, Madingley Road, Cambridge CB3 0HA
\and
Lund Observatory, Department of Astronomy and Theoretical Physics, Box 43, SE-221 00 Lund, Sweden
\and
Instituto de Astrofísica de Andaluc\'ia, Camino Bajo de Huétor, 50, 18008, Granada,  Spain
\and
INAF, Osservatorio Astronomico di Palermo,  Piazza del Parlamento 1, I-90134, Palermo, Italy
\and
Dipartimento di Fisica e Astronomia, Università di Catania, Italy
\and
Nicolaus Copernicus Astronomical Center, Polish Academy of Sciences, ul. Bartycka 18, 00-716, Warsaw, Poland
\and
Instituto de F\'isica y Astronom\'ia, Fac. de Ciencias, Universidad de Valpara\'iso, Gran Bretana 1111, Playa Ancha, Chile
\and
Max-Planck Institut f\"{u}r Astronomie, K\"{o}nigstuhl 17, 69117 Heidelberg, Germany
\and
Dipartimento di Fisica e Astronomia Galileo Galilei, Universitá di Padova, Vicolo Osservatorio 3 I-35122, Padova, Italy
\and
School of Physics and Astronomy, Monash University, Clayton 3800, Victoria, Australia
\and
Faculty of Information Technology, Monash University, Clayton 3800, Victoria, Australia
\and
Departamento de Did\'actica, Universidad de C\'sdiz, 11519 Puerto Real, C\'adiz, Spain
\and
N\'ucleo de Astronom\'ia, Facultad de Ingenier\'ia y Ciencias, Universidad Diego Portales (UDP), Santiago de Chile
\and
Departamento de Ciencias Fisicas, Universidad Andres Bello, Fernandez Concha 700, Las Condes, Santiago, Chile
}

\authorrunning{Hatzidimitriou et al.}
\date{}

  \abstract
      {Pismis~18 is a moderately populated, intermediate-age open cluster located within the solar circle at a Galactocentric distance of about seven kpc. Few open clusters have been studied in detail in the inner disc region before the Gaia-ESO Survey.
      }
      {New data from the Gaia-ESO Survey allowed us to conduct an extended radial velocity membership study as well as spectroscopic metallicity and detailed chemical abundance measurements for this cluster.
      }
      {Gaia-ESO Survey data for 142 potential members, lying on the upper main sequence and on the red clump, yielded radial velocity measurements, which, together with proper motion measurements from the Gaia Second Data Release (Gaia DR2), were used to determine the systemic velocity of the cluster and membership of individual stars. Photometry from Gaia DR2 was used to re-determine cluster parameters based on high confidence member stars only. Cluster abundance measurements of six radial-velocity member stars with UVES high-resolution spectroscopy are presented for 23 elements.
      }
{The average radial velocity of 26 high confidence members is $-27.5\pm2.5(std)~$\kms with an average proper motion of $pmra=-5.65\pm0.08$ (std) ~mas~yr$^{-1}$ and $pmdec=-2.29\pm0.11$ (std)~mas~yr$^{-1}$ . 
According to the new estimates, based on high confidence members, Pismis~18 has an age of   $\tau = 700^{+40}_{-50}$~Myr, interstellar reddening of $E(B-V)=0.562^{+0.012}_{-0.026}$~mag and a de-reddened distance modulus of $DM_0=11.96^{+0.10}_{-0.24}$~mag. The median metallicity of the cluster (using the six UVES stars) is [Fe/H]$=+0.23\pm0.05$ dex, with [$\alpha$/Fe]$=0.07\pm0.13$ and a slight enhancement of s- and r- neutron-capture elements.}
{With the present work, we fully characterized  the open cluster Pismis~18. We confirmed
 its present location in the inner disc. We estimated a younger age than the
 previous literature values  and we gave, for the first time, its metallicity and its detailed abundances.
 Its [$\alpha$/Fe] and [s-process/Fe], both slightly super-solar, are in agreement
with other inner-disc open clusters observed by the Gaia-ESO survey. 
}

\maketitle

\section{Introduction}

Open clusters (OCs), being simple populations with relatively easily determined ages, are among the best tracers of the chemical evolution of the Galactic thin disc from its outer regions to the Galactic bulge \citep[e.g.][]{Friel95,Sestito08,Yong12,Donati15,Magrini15,Cantat-Gaudin16,Netopil16, Casamiquela17}.

The inner disc ($R_{GC}<8$kpc) is an area of particular importance as it constitutes a link between the properties of the bulge and of the thin/thick disc. In order to probe the chemical evolution of the inner disc one needs information  both on ages and on abundances of stars or stellar populations,  which can be provided  by a detailed study of chemical abundances of intermediate age and old clusters with known ages \citep[see, e.g.][]{jacobson16}. Relatively old OCs are quite rare in these high density regions due to high mortality rates \citep{Zwart98,Kruijssen2011}. Thus the percentage of OCs with ages larger than about one Gyr  in the inner disc is only $\simeq$11\%, based on the catalogue of inner disc clusters by \citet{Morales13}. Moreover, inner disc clusters are often difficult to observe due to high field contamination and heavy and/or differential reddening.

Pismis~18 (IC 4291) is considered to be an intermediate age OC with an age of about one Gyr, located in the inner disc, at a distance of about 2.77 kpc (according to a new determination by \citet{cantat2018b} based on Gaia DR2  parallaxes and proper motions) from the Sun in the direction of the Galactic Centre.  It lies within the fourth quadrant of the Galactic plane  on the Sagittarius arm. In the study of \citet{Morales13} it is classified as a totally exposed cluster with no correlation with sub-millimetre dust continuum emission.

Pismis~18 is included in the Gaia-ESO survey (GES), which is a large, public spectroscopic survey of the Galaxy carried out with the high-resolution multi-object spectrograph FLAMES (Fiber Large Array Multi-Element Spectrograph; \citealt{Pasquini02}) on the Very Large Telescope (ESO, Chile). The aim of the survey is to provide accurate radial velocities and detailed element abundances for about $10^5$ stars covering the bulge, thick and thin discs, and halo components, as well as a sample of about 65 OCs of all ages, metallicities, locations, and masses \citep[cf.][]{Gilmore12, Randich13}.
Pismis~18 is the seventh inner disc open cluster individually studied within the framework of the GES, the other clusters being NGC~4815 \citep{Friel14}, NGC~6705 \citep{CantatGaudin14}, Be~81 \citep{Magrini15}, Tr~20 \citep{Donati14}, Tr~23 \citep{Overbeek17} and
NGC~6802 \citep{tang17}. Other studies have focused more on the general properties of the cluster population as a whole \citep[see, e.g.][]{Magrini15, Magrini17, spina17, Bravi18, Randich18}. 

The purpose of this paper is to present our GES observations of Pismis~18, with the aim of providing a detailed membership analysis,  abundances for 23 elements  and revised cluster parameters (age, distance, metallicity and reddening) based on high confidence members. Previous studies of Pismis~18 are summarized in Sect.~2.  In Sect.~3 we describe the target selection, observations and data analysis, while in Sect.~4 we perform the selection of high confidence members based on their proper motions (from the Gaia Second Data Release; hereafter, Gaia DR2, \citealt{gaiacoll2018b}) and GES radial velocities. Sect.~5 provides atmospheric parameters for the high confidence member stars. 
Based on Gaia DR2 photometry of the high confidence cluster members, along with the newly determined metal abundance of the cluster, a revised set of cluster parameters is determined in Sect.~6, and the element abundance measurements are discussed in Sect.~7. Our results are summarized in Sect.~8.

\begin{table*}[t]
\caption{Pismis 18 basic parameters}
\label{t_1}
\setlength{\tabcolsep}{0.05in}
\centering
\small
\begin{tabular}{ l c c c c c c}
\hline\hline
Property & WEBDA &Piatti et al. (1998)	&	Tadross (2008)	&	Kharchenko et al. (2013)&  Cantat-Gaudin et al.(2018a) & Present study\\
\hline
RA (J2000)&13:36:55&13:36:32\tablefootmark{a} &13:36:56&   13:36:55.8       & 13:36:54.5 & 13:36:58.1\\
Dec (J2000)&-62:05:36 &-62:12:48\tablefootmark{a} &-62:05:45& -62:03:54     & -62:05:28   & -62:05:35\\
D (kpc) &2.24&2.24$\pm$0.41&1.79$\pm$0.08&        2.3       &2.77\tablefootmark{b} &$2.47^{+0.11}_{-0.26}$\\
$R_{GC}$ (kpc) & & $-$ &7.53&                  $-$  & &  6.8$\pm$0.1 \tablefootmark{d}   \\
z (pc)\tablefootmark{c}  &        &     &    &                   &   15.1  &  $13\pm2$\\
radius (arcmin)  && $-$ &5.6&                 $-$   & 2.88\tablefootmark{e}   &$\simeq$ 5         \\
age (Gyr)  &&1.2$\pm$0.4&0.8&  0.94               &  &  $0.70^{+0.04}_{-0.05}$\\
E(B-V)  &0.50&0.50$\pm$0.05&0.61&                 0.52  & &  $0.562^{+0.012}_{-0.026}$      \\
{[Fe/H]} &&0.0& $-$ &                         $-$   & &  0.23$\pm$0.05 dex         \\
\hline
\end{tabular}
\tablefoot{\tablefoottext{a}{Converted from the B1950 coordinates given in the paper.}
\tablefoottext{b}{Most likely distance.}
\tablefoottext{c}{Distance from Galactic midplane.}
\tablefoottext{d}{The adopted distance of the Sun to the Galactic Centre is eight kpc \citep[see][]{Malkin2013}.}
\tablefoottext{e}{Radius including 50\% of the cluster stars. }}
\end{table*}

\section{Pismis\,18 in the literature}

The only published optical photometry for Pismis~18 is the CCD $BVI$ photometry of \citet{Piatti98}, who also obtained an integrated spectrum for this cluster. Their colour-magnitude diagram (CMD) reveals a well-defined main sequence (MS) and red clump (RC) with relatively little contamination by field stars, partly due to the small field of view (4$\times$4 arcmin$^2$).   They have estimated the age of Pismis~18 at 1.2$\pm$0.4 Gyr (on the basis of the CMD and the integrated spectrum), with a reddening value of $E(B-V)=0.50\pm0.05$ and a distance of $2.24\pm0.41$ kpc, assuming solar abundance. More recently, \citet{Tadross08} has analysed 2MASS data to construct $JHK$ CMDs.  He has estimated an age of 0.8 Gyr (using solar metallicity isochrones from \citealt{Bonatto04}), interstellar reddening of $E(B-V) = 0.61$ and a distance of 1790$\pm$82pc. He has also provided
revised values for the cluster central coordinates and extent, giving a diameter of 5.6 arcmin, larger than the one tabulated in \citet{Dias02}. The estimates provided by the two studies for the age, reddening and distance of the cluster are marginally consistent within the errors (see Sect.~6). \citet{Kharchenko13} have redetermined the cluster parameters using 2MASS photometry, proper motions, and solar metallicity isochrones. There is no spectroscopic determination of the metallicity of Pismis~18 based on individual stars, although \cite{Piatti98} obtained a rough estimate of the  metallicity, [Fe/H]=0.0,  from the width of CaII triplet lines in  integrated spectra.

Pismis~18 is too distant to be included in the Gaia-TGAS catalogue \citep{Gaia17,Cantat2018a}. However, as mentioned in the Introduction, it is featured in Gaia DR2 \citep[see e.g.][]{cantat2018b} and we use the information in the present paper (relevant parameters from this study are shown in Col. 6 of Table~\ref{t_1}). Table~\ref{t_1} summarizes the literature values of the Pismis~18 parameters, together with the ones derived in the present paper (described in Sect. 6).

Finally, 16 stars in Pismis~18 have been observed by \citet{Carlberg14} with the MIKE spectrograph at a resolution of $\sim 44000$. These spectra are part of a study of rotation in open clusters, so their S/N is low and only permitted derivation of radial velocities (RV) and $v \sin i$. Twelve stars have been classified as members of the cluster with $\langle{\rm RV}\rangle = -27.9 \pm 0.8$ km s$^{-1}$. We have seven stars in common with this study, indicated in Col. 18 of Table~\ref{t_2}.

\section{Observations and data reduction}

The target selection, observation, data reduction, atmospheric parameter determination and abundance measurements were handled within the GES collaboration by specific working groups (WGs). The targets in Pismis~18 were selected following the strategy applied for all intermediate-age OCs with prominent red clumps (RC) \citep[see, e.g.][]{bragaglia2018}.
Briefly, the RC stars, which have high cluster membership probability, were observed with UVES with the aim of determining precise chemical abundances. Stars on the MS were observed with GIRAFFE, using the HR9B setup primarily for stars of spectral type A to F and the HR15N setup for cooler stars (see below for more details). The selection of GIRAFFE targets was aimed at observing an inclusive and unbiased sample of cluster star candidates rather than only high probability members, with the purpose of defining cluster membership using the RVs obtained with the larger GIRAFFE sample.

The MS targets were selected from the MS turnoff down to $V\simeq19$. The GES observations nicely complement with RV the precise astrometric information of the Gaia mission for stars
not reached by the Gaia RVS instrument (Gaia DR2 has RVs only for stars brighter than  G =12, \citealt{Katz18}) and permit a complete chemical characterisation of the cluster with the giants observed with UVES.

\begin{figure*}
\centering
\includegraphics[scale=0.8]{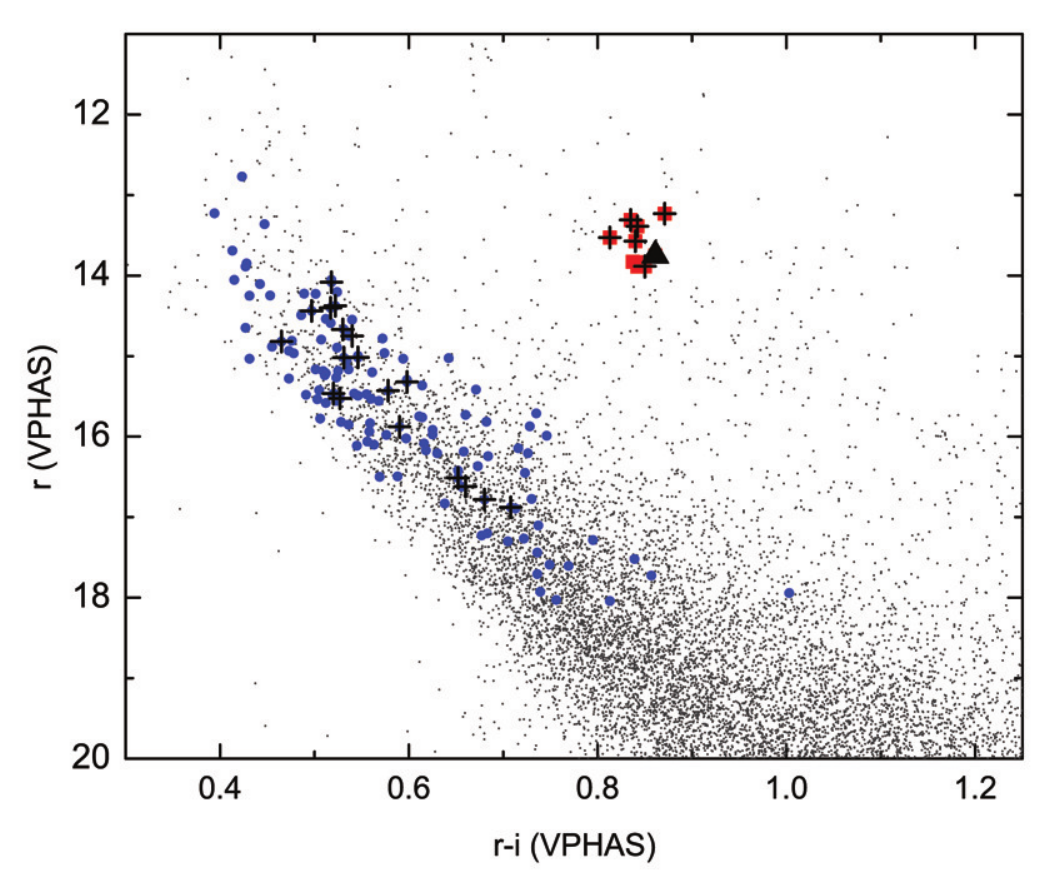}
\includegraphics[scale=0.8]{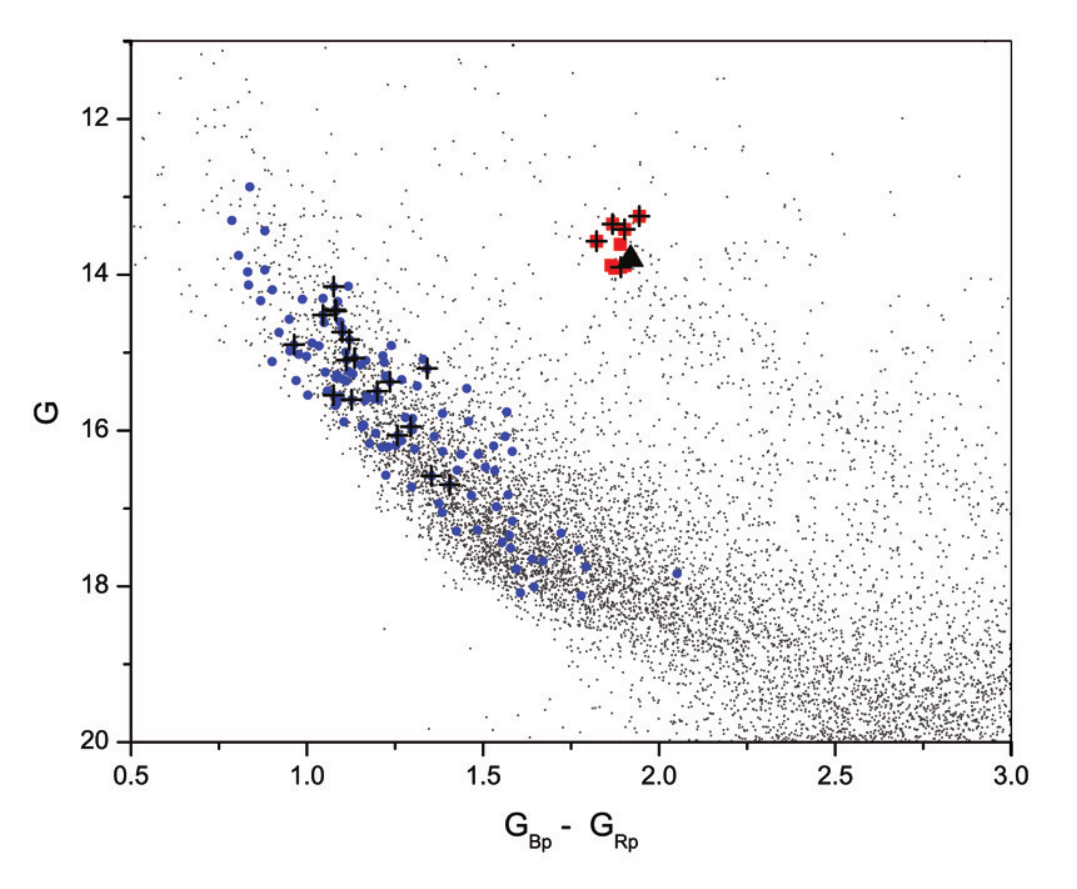}
\caption{{\it Left panel:} VPHAS$+$ colour magnitude diagram of Pismis~18 over the entire field of view of FLAMES, marked as small grey filled circles. The targets observed with UVES are marked with red filled squares, while targets observed with GIRAFFE are marked with blue filled circles. High confidence cluster members are indicated with black crosses and the possible binary star with a black filled triangle;  {\it Right panel:} same, but for Gaia DR2 photometry.}
\label{f_1}
\end{figure*}

The targets were selected on the basis of the VPHAS$+$ ESO survey data \citep{Drew14} in the $r$ and $i-$bands (Vega system), as VPHAS$+$ provides homogeneous spatial coverage over the entire extent of the cluster. In total, we observed ten stars in the RC region and 132 stars on the MS.
The left panel of Fig.~\ref{f_1} displays the $r$ vs ($r-i$) CMD over a region of 15 arcmin around the centre of Pismis~18.  We also mark the selected targets on the MS and the RC. On the right panel of Fig.~\ref{f_1}, we show a similar diagram based on Gaia DR2 photometry. It must be emphasized that not all targets (selected on the basis of VPHAS$+$) have Gaia DR2 photometry in all three bandpasses. It is noted that on the  diagram on the right the main sequence is somewhat narrower for fainter magnitudes, although a direct comparison is not possible due to the different filters used.

The data for Pismis~18 were obtained in May and June 2014  with FLAMES on the VLT-UT2 telescope at the European Southern Observatory. The ten candidate RC stars were observed with the high-resolution spectrograph UVES (Ultraviolet and Visual Echelle Spectrograph, \citealt{Pasquini02}) using the U580 setup (4800-6800 \A\ and $R=47000$), with total exposure times of either 12ks or 15ks.

Spectra were obtained with the medium-resolution multi-fibre spectrograph GIRAFFE for 51 MS stars with $13 <  V < 16$ ($12.7<r< 15.6$) using the HR9B setup ($5143-5356$ \A\ and $R = 25900$), while 91 MS stars with $15.5 < V < 19$ ($15.1 < r < 18.5$ ) were observed using the HR15N setup ($6470-6790$ \A\ and $R = 17000$). It is noted that ten stars were in common between the two configurations. The total exposure times were 15ks for the HR15N setup and 12ks for the HR9B setup. The median signal-to-noise ratio was 116, 74 and 52 for the UVES, GIRAFFE HR15N and GIRAFFE HR9B spectra, respectively.

Data reduction included sky subtraction, barycentric correction and normalisation, as well as calculation of  radial and  rotational velocities.
For details on the data reduction pipeline, specifically for the UVES spectra, see \citet{Sacco14}.

Parameter and abundance determinations for each target were typically performed by multiple nodes (WG sub-groups) in charge of abundance analysis. The results of individual nodes were combined within each WG; the WG values were then homogenized by WG15 (Hourihane et al., in preparation) to yield final recommended parameters, using a set of calibrators to define a common scale \citep{Pancino17}. This structure produced homogeneous parameter determinations while allowing WGs to specialize in different types of stars. More details can be found, for example, in \citet{Overbeek17} and references therein. The data described here came from the fifth internal data release (GES i{\sc dr5}) which comprises a re-analysis of all available spectra until December 2015 using an updated linelist \citep{Heiter15} and state-of-the-art analysis techniques.

In Table~\ref{t_2}, available only electronically in its entirety, we provide for the 142 observed stars,  the GES ID number (CNAME), the Gaia DR2 ID, the equatorial coordinates (J2000) in degrees, the setup used for the observations, the derived radial velocities, the Gaia DR2 magnitudes $Bp$, $G$ and $Rp$, the  $gri$ magnitudes from VPHAS$+$, the 2MASS $JHK$ magnitudes, the WEBDA (a site devoted to Stellar Clusters in the Galaxy and the Magellanic Clouds, developed and maintained by Ernst Paunzen and Christian St\"utz, Institute of Astronomy of the University of Vienna (Austria)\footnote{https://www.univie.ac.at/webda/})
identification number whenever available, the  \citet{Piatti98} $BVI$ photometry,  the angular distance from the cluster centre ($r_{centre}$) (in arcmin), the radial velocity from \citet{Carlberg14} for the common stars, and the parallax (in mas) and proper motion (in mas~yr$^{-1}$) according to Gaia DR2. The label 'm' indicates high confidence membership assigned according to the analysis described in the next section.

\section{Membership determination}

As field contamination is quite significant in OCs, membership determination using radial velocities and proper motions is very important for the derivation of high confidence values for the cluster parameters. Accurate proper motions are now available from Gaia DR2 \citep{Gaia16,gaiacoll2018b}.

\subsection{Proper motions}
\begin{figure}
\centering
\includegraphics[width=0.85\hsize]{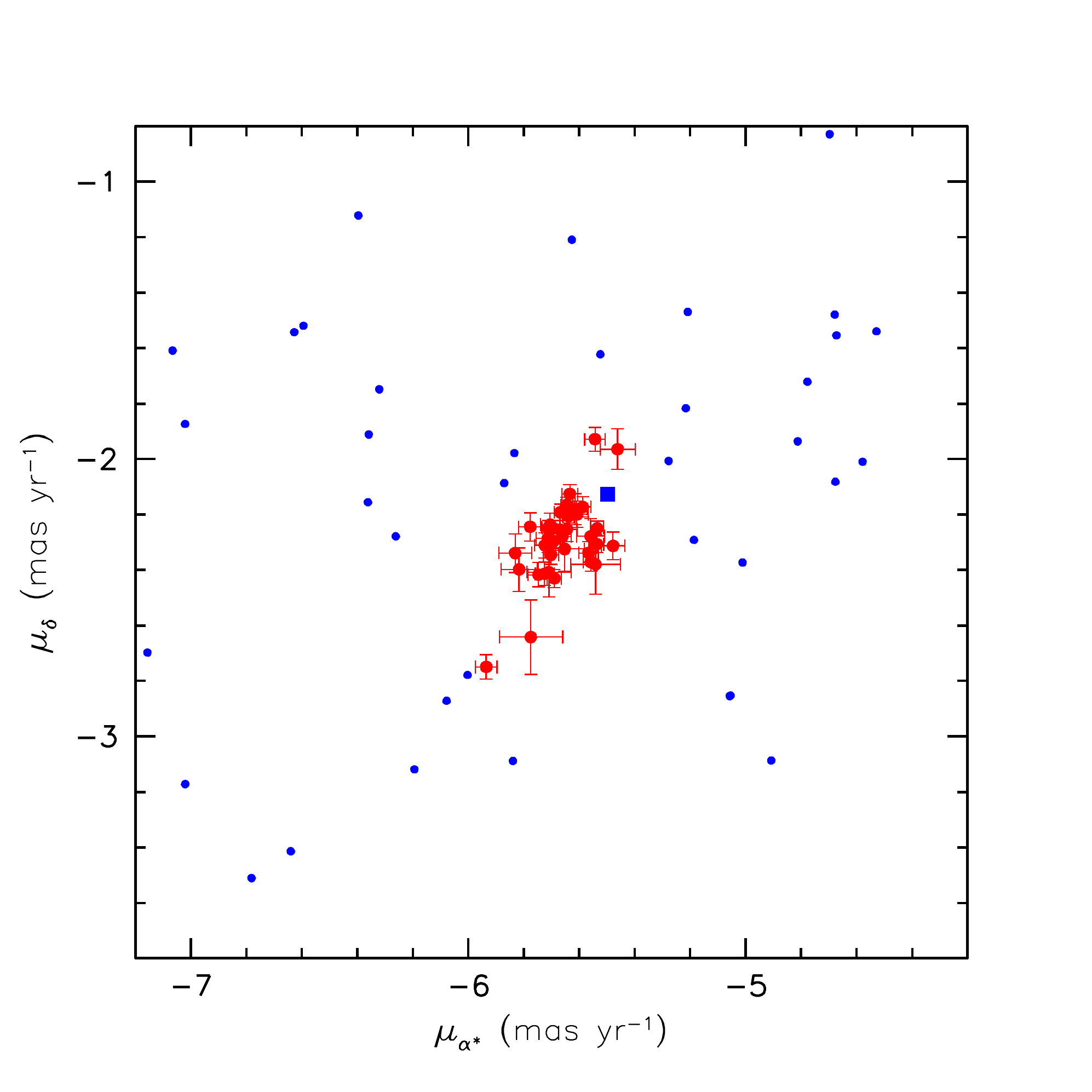}
\caption{
  Proper motions of stars in our Pismis\,18 spectroscopic sample in common with the Gaia DR2.  Red dots represent the probable members according to the analysis of \citet{cantat2018b}, with proper motion errors in the Gaia DR2. A blue square marks a suspect binary star observed with UVES (CNAME 13365001-6205376, see discussion in the text).  Only a limited range of proper motions  is shown here to highlight the locus of cluster member candidates.}

\label{f_pmsel}
\end{figure}

The Gaia DR2 recently provided useful information for the assessment of cluster membership of the targets in our sample.
 A recent study by \citet{cantat2018b} analysed the Gaia DR2 catalogues to derive the membership of stars in a large sample of open clusters, including Pismis\,18. They employed the unsupervised membership assignment code UPMASK  \citep{krone-martins2014} to give a membership probability to each star from proper motions and parallaxes, by taking into account all errors and correlations between the parameters. 
  We took advantage of this analysis to identify a set of probable cluster members among our observed stars. To this aim, our target list was matched with the Gaia DR2 catalogue using the  2MASS identifiers and the pre-computed crossmatch between Gaia DR2 and 2MASS \mbox{("gaiadr2.tmass\_best\_neighbour")}. The latter was preferred as the result of  a careful analysis by the Gaia team, including proper motion propagation and epoch correction \citep{Marrese18}. Only stars with a 5-parameter solution were considered. We found 132 stars in common with Gaia DR2, whose proper motions are plotted in Fig.~\ref{f_pmsel} (blue dots).
  Then, our catalogue was matched with the results of \citet{cantat2018b} using the unique Gaia DR2 identifiers.
We selected the spectroscopic targets with membership probability $P>0.5$ to define a sample of 35 probable members based on astrometry. These are plotted as red dots with error bars in Fig.~\ref{f_pmsel}. Using this sample we computed the average value and standard deviation of proper motions,
$\mu_{\alpha *} = -5.66\pm 0.10 (std)$)~mas~yr$^{-1}$    and
$\mu_\delta = -2.29\pm 0.15 (std)$)~mas~yr$^{-1}$.

The selection of probable cluster members is refined in the next sub-section using the radial velocity information. The proper motions have not been corrected for the systematic uncertainty of the order 0.035  mas~yr$^{-1}$ found by \citet{gaiacoll2018d}.

The Gaia DR2 also provides parallaxes for our 35 cluster member candidates. The mean parallax is quite well defined at
 0.335 $\pm 0.054 (std)$ mas.
However, we refrained from using this value to compute a geometric distance to Pismis\,18 since there is evidence from previous studies that parallaxes in Gaia DR2 are affected by systematic errors of the order 0.03 to 0.05~mas in the parallax absolute zero point  \citep{lindegren2018,luri2018}.  The distance to the cluster is discussed in Sect.~\ref{cl_params}.

 We mention here a special case represented by a UVES star (CNAME=13365001-6205376, marked with a square in Fig.~\ref{f_pmsel}), which has  a proper motion close to that of the probable members, yet it is not present in the \citet{cantat2018b} list. This star might possibly be an unresolved binary, in which case its parallax and proper motion as well as radial velocity could be incorrect. In fact, in Gaia DR2 all sources were treated as single stars in deriving the astrometric solution \citep{lindegren2018,gaiacoll2018b}.
Although not considered as a probable member in the following analysis, this star is included in our tables for future reference as a possible cluster member.

\begin{sidewaystable}
\caption{Photometry and radial velocities for all observed targets.}
\label{t_2}
\setlength{\tabcolsep}{1mm}
\fontsize{6}{8}\selectfont
\begin{tabular}{lcccccccccccccccccccccccc}
\hline\hline
CNAME & Gaia DR2 ID & &RA   & Dec &Setup & RV$^{(1)}$ &BP &$G$ &RP & $g$ & $r$ & $i$ & $J$ & $H$ & $K$ & ID$^{(2)}$ & $V$ & $B-V$ & $V-I$ & $r_{centre}$$^{(3)}$ &RV$_c$$^{(1)}$&parallax$^{(4)}$&pmra$^{(5)}$&pmdec$^{(5)}$\\
   & & &(deg) &(deg) & & & \multicolumn{3}{c}{Gaia DR2}& \multicolumn{3}{c}{VPHAS+} &\multicolumn{3}{c}{2MASS} & & & & & & & & &\\
\hline
13364831-6206517 & 5865731192321534848 &	m     	&	204.2012917	&	-62.1143611	&	 U580 	&	-26.78	&14.543  &13.610  &   12.653		& &	13.574	&	12.734	&	11.156	&	10.515	&	10.274	&	44	&	14.24	&	1.529	&	1.723	&	1.48	&	-27.0	&	0.363	&	-5.668	&	-2.192	\\
13365001-6205376$^{(6)}$& 5865734142929735424	&	m 	&	204.208375	&	-62.0937778	&	 U580 	&	-18.39	&14.704 & 13.800     &12.785	&15.336	&	13.75	&	12.889	&	11.061	&	10.548	&	10.144	&	86	&	14.395	&	1.509	&	1.779	&	0.58	&	-27.3	&	0.492	&	-5.497	&	-2.128	\\
13365321-6210050& 5865729989730573696	&	     	&	204.2217083	&	-62.1680555	&	 U580 	&	-40.28	&14.789    &13.877 &    12.926 &15.377	&	13.833	&	12.99	&	11.408	&	10.733	&	10.532	&	      	&	       	&	       	&	      	&	4.49	&	  	&	0.351	&	-8.204	&	-2.633	\\
13365597-6205130& 5865734181618835456	&	m     	&	204.2332083	&	-62.0869444	&	 U580 	&	-28.13	&14.356  &  13.416 &    12.454 &	14.978	&	13.387	&	12.545	&	10.968	&	10.248	&	10.02	&	26	&	14.035	&	1.571	&	1.735	&	0.4	&	-27.8	&	0.320	&	-5.711	&	-2.251	\\
13365882-6205197& 5865731226681330560	&	m     	&	204.2450833	&	-62.0888055	&	 U580 	&	-27.38	&14.832  & 13.903  &   12.941&	15.451	&	13.884	&	13.034	&	11.449	&	10.746	&	10.556	&	22	&	14.546	&	1.538	&	1.745	&	0.52	&	-27.3	&	0.324	&	-5.690	&	-2.430	\\
13370214-6206095& 5865731157961823360 	&	m     	&	204.2589167	&	-62.1026389	&	 U580 	&	-26.82	&14.460     &  13.570 &     12.638	&15.048	&	13.527	&	12.714	&	11.084	&	10.548	&	10.306	&	18	&	14.161	&	1.472	&	1.675	&	1.01	&	-26.5	&	0.319	&	-5.645	&	-2.165	\\
13370523-6206433& 5865731325431190400	&	m     	&	204.2717917	&	-62.1120278	&	 U580 	&	-27.79	&14.268  & 13.350 &     12.400 &	14.877	&	13.31	&	12.475	&	10.931	&	10.252	&	10.02	&	9	&	13.977	&	1.527	&	1.724	&	1.64	&	-27.1	&	0.350	&	-5.535	&	-2.250	\\
13371182-6206030& 5865731364120257024	&	m     	&	204.29925	&	-62.1008333	&	 U580 	&	-28.32	&14.214        & 13.247    &12.270	&14.856	&	13.231	&	12.36	&	10.737	&	10.036	&	9.779	&	2	&	13.907	&	1.593	&	1.783	&	2.02	&	-29.4	&	0.301	&	-5.639	&	-2.206	\\
13372623-6204585& 5865731501559248768 	&	     	&	204.3592917	&	-62.0829167	&	 U580 	&	 -5.87	&14.825            & 13.881    & 12.922 &	15.476	&	13.828	&	12.991	&	11.417	&	10.711	&	10.444	&	      	&	       	&	       	&	      	&		&	3.71	&	0.349	&	-11.095	&	-2.397	\\
13372876-6202527& 5865737651952931328	&	     	&	204.3698333	&	-62.0479722	&	 U580 	&	-33.90	&14.841             &  13.918 &     12.967	& 15.445	&	13.888	&	13.046	&	11.483	&	10.787	&	10.585	&	      	&	       	&	       	&	      	&		&	4.8	&	0.395	&	-3.964	&	-1.979  	\\
\hline
\end{tabular}
\\
\tablefoot{The full table is available at the CDS. Column 22 gives the radial velocity determinations of \citet{Carlberg14}. The letter 'm' (Col. 3) indicates high confidence membership, according to the analysis described in Sect.~4. Errors are provided (whenever available) for the  magnitudes, radial velocities, parallaxes and proper motions in the electronic version of the table.
$^{(1)}$ In \kms;
$^{(2)}$~WEBDA numbering system; 
$^{(3)}$ in arcmin; 
$^{(4)}$ in mas; 
$^{(5)}$ in mas~yr$^{-1}$;   
$^{(6)}$ possible binary star in Pismis\,18.
} \\
\end{sidewaystable}

\begin{figure}
\centering
\includegraphics[width=8cm,scale=1.0]{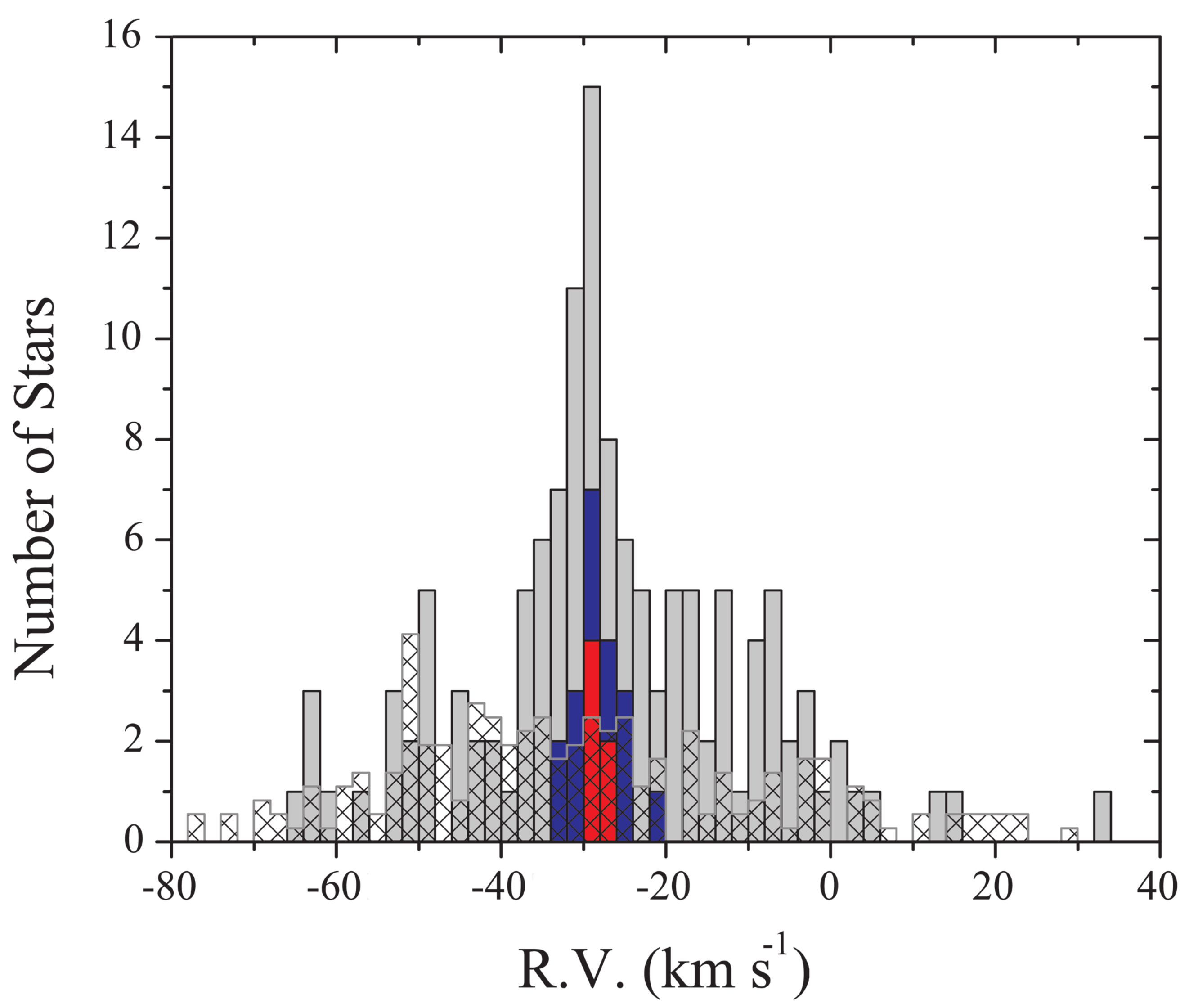}
\caption{Distribution of radial velocities of all 142 observed stars (in grey). The distribution of the GIRAFFE radial velocity members is shown in blue and that of the UVES radial velocity members in red.
The patterned bars show the distribution of radial velocities of Galactic stars in the direction of Pismis~18 according to the Besan\c{c}on star count model for the Galaxy. For the normalization of this curve see text. }
\label{f_2}
\end{figure}

\subsection{Radial velocities}

The 35 proper motion likely members selected in the previous sub-section, were further analysed on the basis of their radial velocities, in order to construct the final catalogue of high confidence members based on both proper motions and radial velocities. The entire sample of the 142 observed Pismis~18 targets, given in Table~\ref{t_2}, shows a wide range of radial velocities (from $-63$ to $+186$~\kms) with a broad peak around $-24$~\kms with a standard deviation of 26~\kms (shown in grey in Fig.~\ref{f_2}). 
The 35 proper motion likely members have a much tighter radial velocity distribution (from $-61.7$  to $-13.9$~\kms), with an average  of $-28.0$ and a standard deviation of 7.4~\kms.
It must be noted that the radial velocity errors for individual stars are quite low, specifically, the median error was $\sim 0.51$~\kms for HR15N, 0.54~\kms for HR9B and 0.36 \kms for UVES (these values refer to the entire sample of observed targets). Stars with high rotational velocities or low signal-to-noise ratios can have significantly higher errors (several \kms, cf. \citealt{Jackson15}). We have thus excluded from further analysis stars with radial velocity errors larger than 5~\kms. 

For the remaining 32 proper motion likely members with radial velocity errors less than 5~\kms, we applied an iterative 2$\sigma$ clipping procedure on the mean, until no stars could be eliminated as outliers. This was achieved in just four iterations, providing a catalogue of  26 stars, which are considered to be high confidence cluster members.  Based on this sample of high confidence members, the average radial velocity of Pismis~18 becomes  -27.5~\kms with a standard deviation of 2.5~\kms and a median of -27.8~\kms. It is noted that no systematic differences were found between radial velocities derived with different setups, within the corresponding errors. 

Among the  26 high confidence members, there are six stars observed with UVES, for which there are also detailed metal abundances, discussed in Sect.~7. Radial velocities for these six stars have also been obtained by \citet{Carlberg14} (see Col. 18 of Table~\ref{t_2}). These values are in good agreement with our measurements within the quoted errors.  A seventh UVES star in common with \citet{Carlberg14}, CNAME13365001-6205376, also mentioned in Sect.~4.1, shows a discrepancy since it has $RV = -27.3$~\kms in \citet{Carlberg14} and $-18.4$~\kms in our study. This discrepancy could be accounted for by the possible binary nature of this object. Although it might be a cluster member, it has not been included in the analysis of high confidence members.

In Fig.~\ref{f_2} we show the distribution of the radial velocities of the  20 GIRAFFE and six UVES high confidence member stars (blue and red histogram, respectively), as well as the radial velocity distribution of all 142 observed stars (indicated in grey). The  patterned bars correspond to the distribution of the radial velocities of Galactic stars in the direction of Pismis~18 according to the Besan\c{c}on star count model for the Galaxy (http://model.obs-besancon.fr/) (\citealt{Robin03}). The predicted distribution is scaled to the total number of observed stars divided by the number of stars in the VPHAS$+$ photometry, in the same $V$ mag range and the same sky area. The scaled model predicts that $\simeq$~60 field stars fall within the selected boundaries of the cluster area, while 6-7 of these fall within the radial velocity range of the Pismis~18 members.

\begin{figure}
\centering
\includegraphics[height=7cm,scale=1]{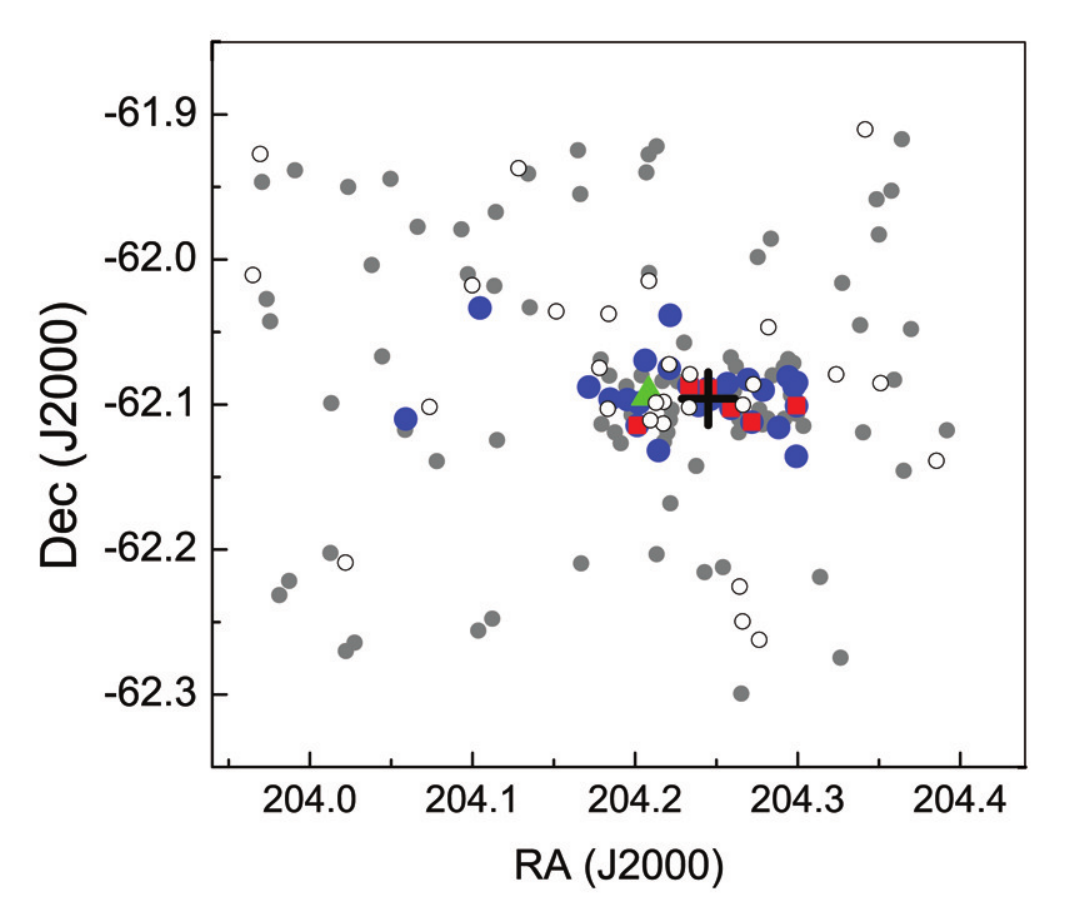}
\caption{Distribution of stars with assigned cluster membership, on the RA-Dec plane. With small grey filled circles we indicate all observed stars, with large blue filled circles the GIRAFFE radial velocity members, with red filled squares the UVES radial velocity members, with a green filled triangle the possible binary star discussed in the text and with a black cross we indicate the location of the cluster centre (as determined in Sect.~6). With open grey circles we indicate stars with RVs consistent with cluster membership, although their proper motions lie outside the 3$-\sigma$ acceptance radius discussed in Sect.~4.1 and shown in Fig.~\ref{f_pmsel} }
\label{f_3}

\end{figure}

Figure~\ref{f_3} shows the location of the 26 radial velocity members on the RA-DEC plane. The stars assigned cluster membership according to their proper motion and radial velocity show a clear concentration close to the cluster centre, with the majority (24 of the  26) lying within a radius of 3.3 arcmin. The remaining two stars lie at distances up to $\simeq $ 5 arcmin (3.6pc for the estimated distance of the cluster, see Sect.~6) from the cluster centre. Although the list of cluster members considered here is not exhaustive, as only high confidence members have been included, it seems that the radius of the cluster is close to 5 arcmin, similar to the estimate of \citet{Tadross08}. \citet{cantat2018b} have estimated that the radius enclosing half the high probability members (based on Gaia DR2 proper motions and parallaxes) is 2.88 arcmin. Following a similar approach, we derived a radius 1.48 arcmin enclosing half the high confidence members in our sample. This value is significantly lower than that of \citet{cantat2018b} and  remains unchanged if we consider all 35 proper motion members (i.e. without the RV selection). Apart from the small size of our sample, our targets were selected on the basis of their location on the CMD, without attempting a homogeneous spatial coverage of the field of view. Therefore,  statistically the derivation of the radius is not very meaningful, and comparisons with values derived from larger and homogeneous samples are not informative.

It must be noted that proper motions are of paramount importance in assigning cluster membership, as there are several objects with radial velocities consistent with cluster membership (open grey circles in Fig.~4), but which are generally scattered over larger distances from the cluster centre and have inconsistent proper motion values.
The  average proper motion of the 26 high confidence members of Pismis~18 is $pmra=-5.65\pm0.08$ (std) ~mas~yr$^{-1}$ and $pmdec=-2.29\pm0.11$ (std)~mas~yr$^{-1}$.

To summarize, six out of the ten UVES target stars and 20 out of the 132 GIRAFFE targets are considered as high probability cluster members on the basis of their proper motions, parallaxes and RV.  The calculated cluster radial velocity dispersion of 2.5~\kms is higher than the $\simeq 1$~\kms  expected for a typical OC \citep[e.g.][]{Mermilliod09}. The larger dispersion could be caused by the fact that our sample consists mostly of upper main sequence stars which are more likely to rotate than the solar-type dwarf stars used by \citet[]{Mermilliod09}. If we use only the six UVES high confidence members lying on the red clump, the velocity dispersion falls to 0.7~\kms.

\section{Atmospheric parameters}

The i{\sc dr5} database provides effective temperatures, $T_{\rm eff}$, for a total of 133 of the 142 observed stars, and surface gravities ($\log g$) and metallicities ([Fe/H]) for 123 of them. Metallicities based on GIRAFFE data are highly uncertain and therefore they are not used in our analysis. 
Of the 20 GIRAFFE high confidence radial members,   17 have measured  $T_{\rm eff}$ and 14 of those have $\log g$ measurements. 
We provide the atmospheric parameters for the high probability member stars in Table~\ref{tab_atmpar}; we have kept the probable binary in this table and following ones even if its values were not used to compute cluster averages.
Based on the six UVES members, the median metallicity of the cluster is [Fe/H]$=0.23\pm0.05$~dex. This value was adopted as the cluster metallicity in the cluster parameter determination described in the next section.  It is noted that the metal abundance derived for the likely binary star is entirely consistent with the cluster value.
It is noted that the global metallicity [Fe/H] of Pismis~18 derived in the present work is higher than that  given in previous papers that adopt the results of Gaia-ESO {\sc dr4} \citep[0.11$\pm$0.02 and 0.10$\pm$0.03, respectively, in][]{jacobson16,Magrini17}. The difference is due to the improved process of homogenization and combination of Node results followed in i{\sc dr5}. In particular, during i{\sc dr5} a lower number of Nodes participated in the analysis of the UVES WG11 spectra, compared to {\sc dr4}. Most of the Nodes considered in the homogenization of [Fe/H] for i{\sc dr5} based their derivations on the Equivalent Width (EW) analysis and tend to obtain a slightly higher [Fe/H] than that obtained using spectral synthesis (see \citet{Jofre2017} for a discussion of systematic differences between different methods of metal abundance derivation). 
On the other hand, the [FeI/H] and [FeII/H]  abundances recomputed by Nodes after the homogenization of the stellar parameters is much closer to the results of the  {\sc dr4} release (see Table~8, 0.04$\pm$0.06 and 0.08$\pm$0.07, respectively).

\section{Redetermined cluster parameters} \label{cl_params}

Using our membership assessment, we re-evaluated the fundamental  parameters of Pismis~18. This evaluation has been based on a relatively small number of stars, spanning a relatively limited range in masses (upper main sequence and clump stars). However, as they are high confidence members, they provide a good comparison against other methods, which may be affected by contamination from field stars.

The centre of the cluster was determined to be at RA$=13^{\rm h} 36^{\rm m} 58.1^{\rm s}$, Dec$=-62\degr 05\arcmin 35\arcsec$ from the median values of the coordinates of the 26 high confidence members (Table~\ref{t_1}: all coordinates are J2000).
 The right ascension agrees with all previous studies (including the WEBDA value)  within $\sim 1\arcmin$, except for the study of  \citet{Piatti98} where  a large difference  of about 7$\arcmin$ is noticed. In declination, the agreement with WEBDA and \citet{Tadross08} is excellent (within ten $\arcsec$), while there are discrepancies from two to seven arcmin with the other studies. The good agreement with the results of \citet{Tadross08} in both coordinates is encouraging in view of the different sample and method used.

The stellar population parameters (age, reddening, and de-reddened distance modulus) were derived by comparing Pisa theoretical isochrones (computed on purpose) with the recently made available Gaia DR2 photometry  \citep[]{gaiacoll2018b}, by adopting the same maximum-likelihood technique described in detail in \citet{Randich18}. To show the robustness of the derived parameters we compared the cluster sequence obtained with the best fit set of parameters in several observational planes, using the $BV$-bands from \citet{Piatti98}, the $JHK_s$ (2MASS) and the $gri$ ($VPHAS+$) photometry, along with the $G_\mathrm{Bp}$, $G_\mathrm{Rp}$ and $G$ Gaia magnitudes. All available photometric information is provided in  Table~\ref{t_2} for all 142 observed stars (including the 26 high confidence members). 

We computed stellar models using the Pisa evolutionary code \citep{Deglinnocenti08,Tognelli11,Dellomodarme12}, with the same input physics described in \cite{Randich18} and \citet{Tognelli18}, except for the initial chemical composition, for which we adopted the value obtained from the six cluster members oberved with UVES (see Sect.~5). More specifically, we built a grid of models using three values of [Fe/H], namely the fiducial value [Fe/H]~=~$+0.23$~dex, and the two extremes (lower and upper limit) obtained assuming the estimated uncertainty $\Delta $[Fe/H]~=~$\pm 0.05$~dex. We assumed the solar-scaled metal distribution given by \citet[][hereafter AS09]{Asplund09}, thus [$\alpha$/Fe]~=~0, which is consistent with that measured for the six members observed with UVES (i.e. [$\alpha$/Fe]~=~$+0.07\pm0.13$, see Sect.~7). In addition, we verified that for the adopted metallicity, a change in [$\alpha$/Fe] by $\pm0.1$ has a negligible impact on the derived parameters. Adopting the AS09 solar-scaled metal distribution, a helium-to-metal enrichment ratio of two \citep{Casagrande07} and a primordial helium abundance of $Y_p=0.2485$ \citep{Cyburt04}, we derived an initial helium and metallicity of ($Y$,$Z$) = (0.291,0.0212), (0.287,0.0191), and (0.296,0.0235), for the fiducial, lower and upper [Fe/H] values, respectively. 

\begin{table*}
\caption{Atmospheric parameters for high confidence members.}
\label{tab_atmpar}
\setlength{\tabcolsep}{1.5mm}
\centering
\begin{tabular}{lrrrrrrrrc}
\hline\hline
CNAME & $T_{\rm eff}$ &Err & $\log g$ & Err & [Fe/H] & Err & $\xi$ & Err &Setup$^{(1)}$\\
\hline
13364831-6206517 &4983 &  60 &  2.85 &  0.12 &  0.23 & 0.10 & 1.545 &0.046 &U\\
13365597-6205130 &4882 &  60 &  2.62 &  0.12 &  0.23 & 0.10 & 1.655 &0.127 &U\\
13365882-6205197 &5045 &  60 &  3.01 &  0.12 &  0.14 & 0.10 & 1.695 &0.174 &U\\
13370214-6206095 &4933 &  60 &  2.81 &  0.12 &  0.29 & 0.10 & 1.565 &0.104 &U\\
13370523-6206433 &4861 &  60 &  2.64 &  0.12 &  0.22 & 0.10 & 1.655 &0.156 &U\\
13371182-6206030 &4950 &  60 &  2.74 &  0.11 &  0.21 & 0.10 & 1.760 &0.082 &U\\
13365001-6205376$^{(2)}$ &4955 &  60 &  2.81 & 0.11  &  0.22 & 0.10 & 1.390 &0.026 &U\\
13361412-6206360 &7750 &  51 &       &	     &	     &	    &	    &	   &G\\  			
13362510-6202004 &     &     &	     &	     &	     &	    &	    &	   &G\\  			
13364117-6205166 &8520 & 750 &  3.55 &  0.45 &       &	    &	    &	   &G\\  
13364430-6205471 &7022 & 120 &  4,37 &  0.16 &       &      &       &      &G\\
13364687-6205483 &8981 & 750 &  3.96 &  0.30 &       &	    &	    &	   &G\\  	
13364855-6205555 &7498 &  96 &  4.25 &  0.21 &       &      &	    &	   &G\\  		
13364946-6204100 &6633 & 148 &  4.21 &  0.14 &       &      &	    &	   &G\\  	
13365149-6207542 &     &     &	     &	     &	     &	    &	    &	   &G\\  				
13365304-6204298 &6791 & 102 &  4.23 &  0.13 &       &      &	    &	   &G\\  	
13365318-6202181 &8378 & 750 &  3.7  &  0.45 &       &	    &	    &	   &G\\
13365737-6206023 &8613 & 750 &  3.72 &  0.45 &       &	    &	    &	   &G\\
13365917-6205472 &     &     &	     &	     &	     &	    &	    &	   &G\\  				
13370162-6205086 &7961 &  62 &       &	     &	     &	    &	    &	   &G\\  				
13370473-6204579 &8077 & 750 &  3.5  &  0.45 &       &	    &	    &	   &G\\  	
13370693-6205236 &8801 & 750 &  3.80 &  0.30 &       &	    &	    &	   &G\\  	
13370918-6206569 &8779 & 750 &  4.12 &  0.20 &       &	    &	    &	   &G\\  	
13371063-6204512 &7004 & 500 &  3.72 &	0.45 &       &      &       &      &G\\
13371147-6205141 &7121 &  92 &       &	     &	     &	    &	    &	   &G\\  	
13371180-6208085 &8613 & 750 &  3.63 &  0.45 &       &	    &	    &	   &G\\
13371184-6205052 &7968 & 750 &  3.00 &  0.70 &       &      &       &      &G\\
\hline
\end{tabular}
\tablefoot{$^{(1)}$U:UVES, G: GIRAFFE;   $^{(2)}$possible binary star.}
\end{table*}

For the model computation we used our solar-calibrated mixing length parameter ($\alpha_\mathrm{ML}=2.0$), which was assumed to be the same for stars with different masses and/or in different evolutionary stages. We included a step core overshooting in the models, for 
$M\ge 1.2$~$M_{\odot}$, with a standard value of $\beta_\mathrm{ov}=0.150$ \citep{Valle17}. From the evolutionary tracks we obtained the isochrones in the age range 300~Myr-2~Gyr with a grid spacing of ten Myr, a good compromise between a dense enough and not extremely large grid to achieve a good age resolution.

To derive the cluster parameters, we opted for Gaia DR2 photometry, which has small  uncertainties ($\Delta$m(G)$<0.001$~mag and $\Delta$m(Bp) and
$\Delta$m(Rp)$< 0.01$~mag), thus resulting in well constrained values for these parameters. To properly account for the extinction/reddening in the Gaia bands, we adopted the extinction law given in Eq.(1) in \citet{gaiacoll2018a}. In addition to Gaia DR2 photometry, we also applied the same method using $VPHAS+$ photometry.  The derived parameters are fully compatible with those obtained using the Gaia DR2 photometry. We did not use the \citet{Piatti98} $BVI$ and the 2MASS photometry in the same manner, because, for the former the uncertainties were not available, while for the latter the CMD shows significant scatter leading to large uncertainties for the values of the derived parameters. 

As already mentioned, we adopted the maximum-likelihood technique described in \citet{Randich18}, which they applied to young clusters in the TGAS catalogue. However, instead of assuming a fixed cluster distance  based on TGAS parallaxes as was done in \citet{Randich18}, we treated the cluster distance as a free parameter, as Gaia DR2 parallaxes for relatively distant objects may suffer from non-negligible  systematic errors of the order of 0.05~mas or more \citep[see e.g.][]{cantat2018b,Riess18,stassun2018,Zinn18}. Such a systematic error  would affect significantly the cluster distance and, as a result, the quality of the isochrone fitting, as is further discussed later. 

We recall that in the adopted maximum-likelihood technique the best values of the vector of cluster parameters (age, reddening, distance) = ($\tau$, $E(B-V)$, $DM_0$) are estimated together. To properly evaluate the confidence interval (hereafter CI) of such quantities, we adopted a Monte Carlo procedure. We perturbed independently the photometric data for each star in each band using the available information on the uncertainty (which we assumed to be Gaussian) to obtain $N$ representations of the same cluster. We set $N=100$, which is large enough to guarantee convergence. Thus, for each perturbed sample $j$ we derived the vector ($\tau_j$, $E(B-V)_j$, $DM_{0,j}$)). The best value for each one of the parameters and its CI were obtained from the ordered sample of the $N$ simulations, by taking the mid value of the distribution (best value), and the 16 and 84 percentile (which define the confidence interval, i.e. the uncertainty). This approach could account only for the observational uncertainties on the photometric bands. However, we wanted to give an estimation of the uncertainties in $\tau$, $E(B-V)$ and $DM_0$ due to the adopted chemical composition. To this purpose, we computed models for the upper and lower limit of [Fe/H] and used these two grids to re-derive the cluster parameters and their CI (using Monte Carlo simulations for the photometry). The effect of adopting a different chemical composition was to shift the best values of the derived parameters with respect to that estimated using the fiducial value of [Fe/H]. We assumed this shift to be representative of the errors due to $\Delta$[Fe/H], and incorporated in the errors caused by photometry alone. We found that $\Delta$[Fe/H] accounted for about one half of the uncertainty on the estimated age, reddening and distance modulus.

\begin{figure*}
\centering
\includegraphics[scale=0.33]{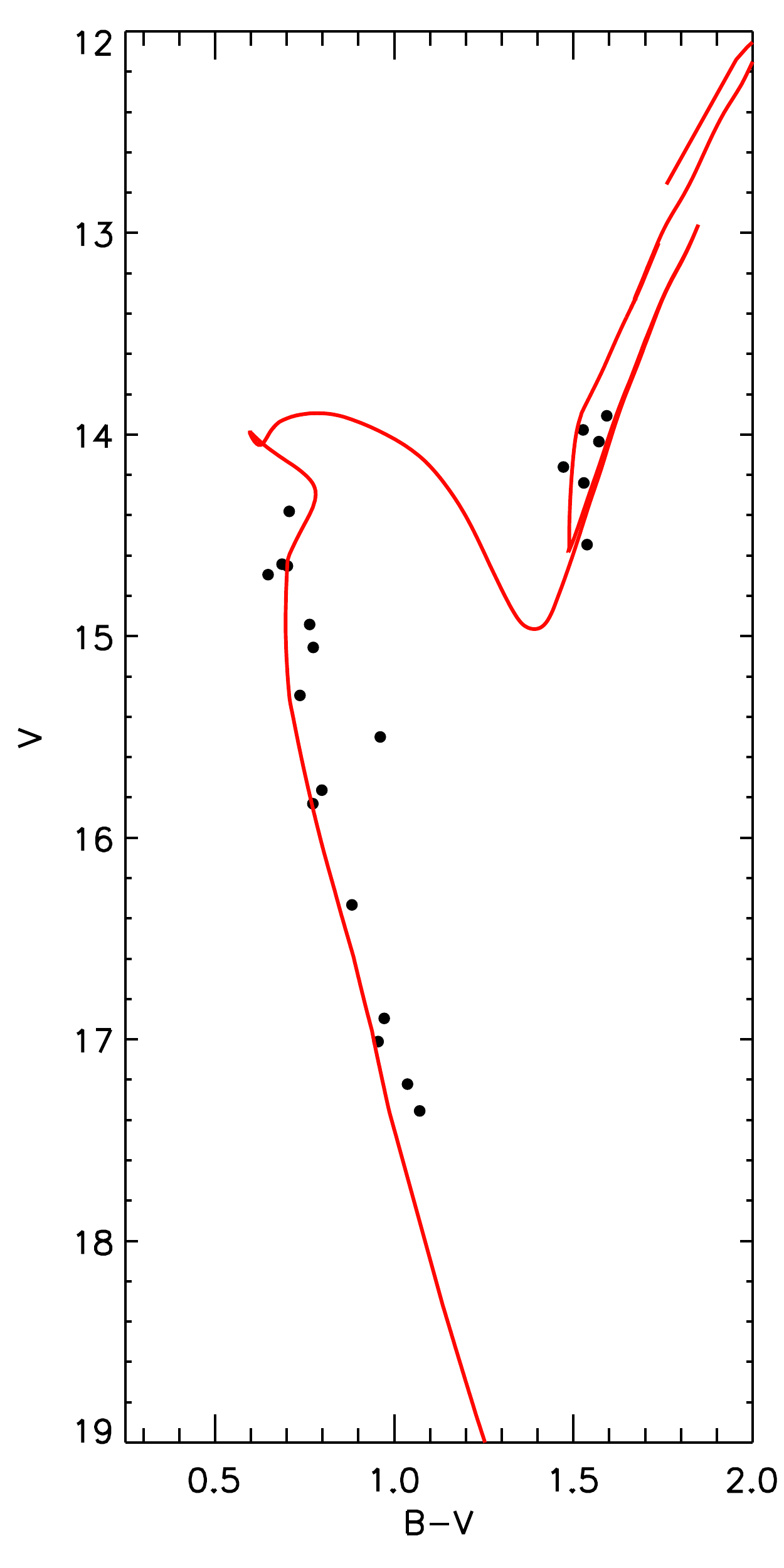}
\includegraphics[scale=0.33]{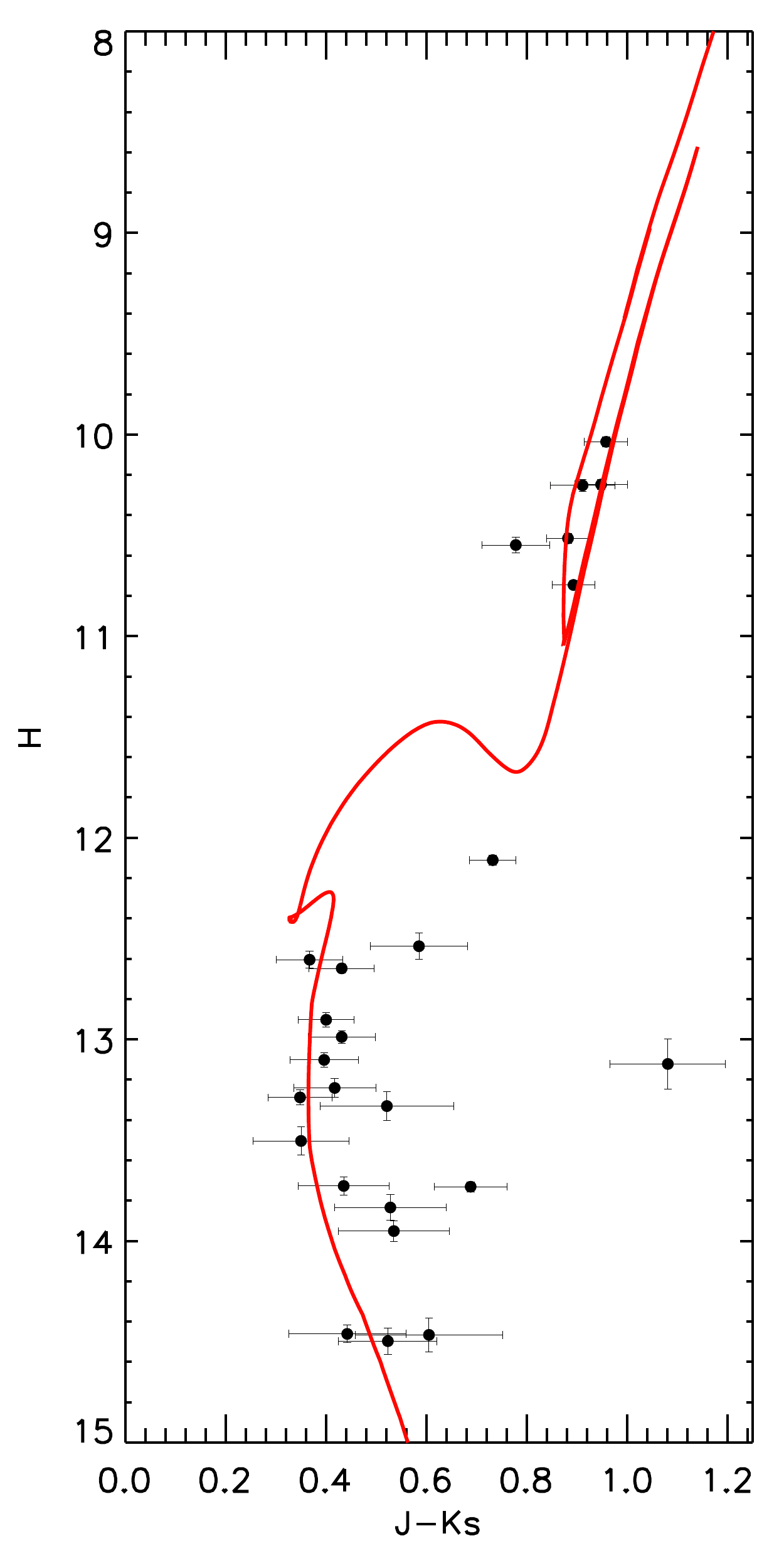}
\includegraphics[scale=0.33]{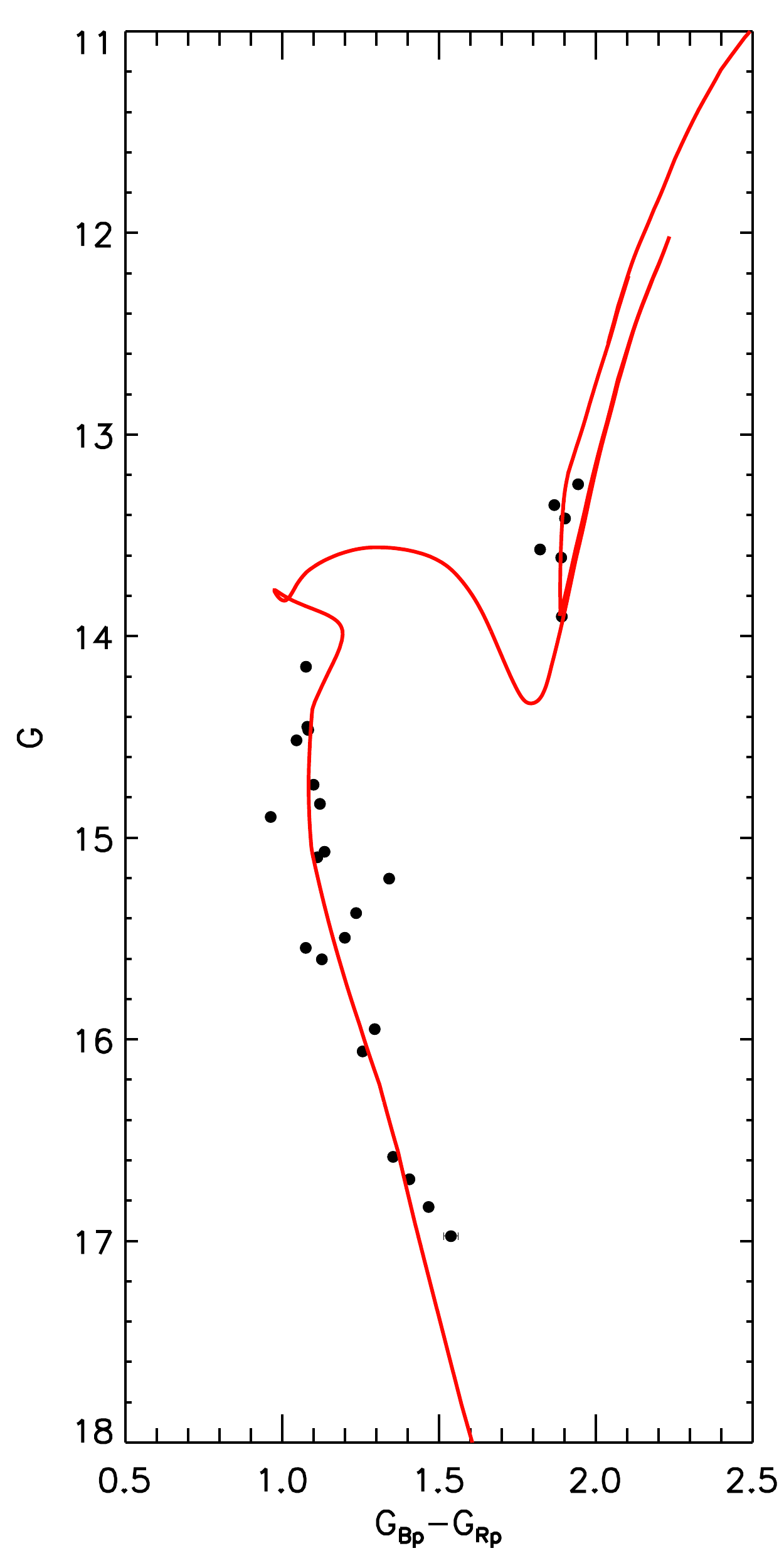}
\includegraphics[scale=0.33]{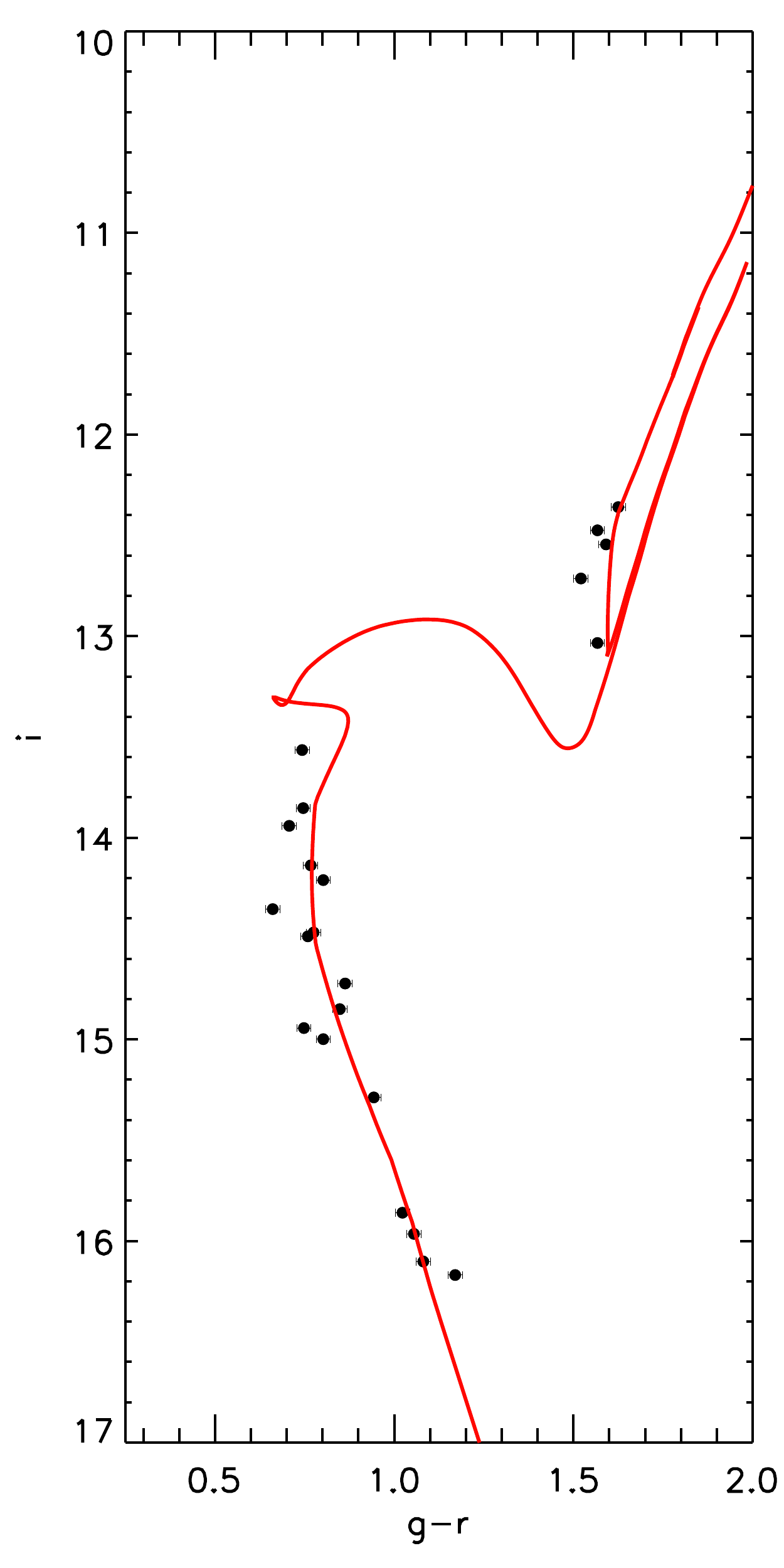}
\caption{Comparison between the Pisa best isochrone (corresponding to $\tau = 700^{+40}_{-50}$~Myr, $E(B-V)=0.562^{+0.012}_{-0.026}$~mag and $DM_0=11.96^{+0.10}_{-0.24}$~mag) and the observations in several photometric bands, namely $BV$ \citep{Piatti98}, $JHK_s$ (2MASS), $G_\mathrm{Bp}G_\mathrm{Rp}G$ (Gaia DR2), and $gri$ ($VPHAS+$). The uncertainties in magnitude and colour are too small to show on the two right-most  diagrams. No uncertainties were available for the BV photometry. }
\label{fig:cmds}
\end{figure*}

We show the comparisons between the best set of isochrones (best $\tau$, $E(B-V$ and $DM_0$) and data in several photometric bands in the panels of Fig.~\ref{fig:cmds}. Our best estimate led to an age of $\tau = 700^{+40}_{-50}$~Myr, a reddening of $E(B-V)=0.562^{+0.012}_{-0.026}$~mag and a de-reddened distance modulus of $DM_0=11.96^{+0.10}_{-0.24}$~mag (i.e. a mean distance of $2.47^{+0.11}_{-0.26}$~kpc and a mean parallax of $0.406^{+0.047}_{-0.019}$~mas).  The quoted distance error encompasses the uncertainties related  to our choice of stellar models. Indeed, a quick independent interactive isochrone fit with different sets of evolutionary models (PARSEC, BaSTI and Dartmouth) confirmed the 
given distance modulus within 0.1 mag. As a general comment, we note that the best isochrone achieved a very good agreement with the data in all the adopted photometric bands for both MS and RC stars, with the exception of the 2MASS CMD which shows large scatter; however, even in this case, the best isochrone could reproduce the RC stars. As discussed in Sect.~5, the metallicity of Pismis~18 according to GES DR4 is significantly lower (around [Fe/H]=+0.10 dex) than the value derived in the present paper.  Adopting this lower metallicity and applying the same procedure, we derived a distance modulus of $DM_0=11.91$ mag, which is 0.05 mag lower than the one obtained for [Fe/H]=+0.23, but within the formal uncertainties. Similarly,  the derived age was reduced by ten Myr, again within the quoted errors. However, the reddening value obtained  was higher by about 0.038 mag, which is about 3$\sigma$ larger than the value for obtained for [Fe/H]=0.23.

\begin{table*}
\caption{UVES members light and $\alpha$-element abundances}
\label{t_5}
\setlength{\tabcolsep}{1mm}
\centering
\footnotesize
\begin{tabular}{cccccccccccccc}
\hline\hline
CNAME&\ion{Li}{I}&\ion{Na}{I}&\ion{Mg}{I}&\ion{Al}{I}&\ion{Si}{I}&\ion{Si}{II}&\ion{Ca}{I}&\ion{Ca}{II}&\ion{Ti}{I}&\ion{Ti}{II} \\
\hline
13364831-6206517&0.7 $\pm$0.05&6.51$\pm$0.05&7.58$\pm$0.14&6.46$\pm$0.06&7.66$\pm$0.08&7.57$\pm$0.13	      &6.48$\pm$0.08&6.22$\pm$0.05&4.92$\pm$0.08&5.08$\pm$0.11\\
13365597-6205130&0.53$\pm$0.07&6.51$\pm$0.05&7.59$\pm$0.12&6.43$\pm$0.08&7.69$\pm$0.07&7.63$\pm$0.14	      &6.44$\pm$0.07&6.47$\pm$0.08&4.87$\pm$0.09&5.08$\pm$0.13\\
13365882-6205197&1.43$\pm$0.06&6.47$\pm$0.06&7.59$\pm$0.17&6.41$\pm$0.08&7.59$\pm$0.09&7.61$\pm$0.16	      &6.42$\pm$0.08&		  &4.91$\pm$0.11&4.96$\pm$0.12\\
13370214-6206095&0.20$\pm$0.30 &6.48$\pm$0.05&7.61$\pm$0.09&6.45$\pm$0.05&7.70$\pm$0.08&7.57$\pm$0.18	      &6.47$\pm$0.07&6.37$\pm$0.09&4.90$\pm$0.08&5.12$\pm$0.15\\
13370523-6206433&1.06$\pm$0.05&6.47$\pm$0.07&7.52$\pm$0.14&6.40$\pm$0.07&7.66$\pm$0.09&7.67$\pm$0.23&6.38$\pm$0.08&		  &4.81$\pm$0.09&5.00$\pm$0.16\\
13371182-6206030&0.58$\pm$0.07&6.53$\pm$0.10&7.60$\pm$0.12&6.52$\pm$0.08&7.69$\pm$0.09&7.51$\pm$0.03&	      6.46$\pm$0.07&6.12$\pm$0.09&4.96$\pm$0.09&5.08$\pm$0.14\\
13365001-6205376&1.34$\pm$0.05&6.48$\pm$0.07&7.58$\pm$0.09&6.44$\pm$0.05&7.67$\pm$0.08&7.64$\pm$0.08&	      6.49$\pm$0.06&6.26$\pm$0.01&4.93$\pm$0.09&5.09$\pm$0.11\\
\hline
\end{tabular}
\end{table*}

\begin{table*}
\caption{Iron-peak abundances for UVES members}
\label{t_6}
\setlength{\tabcolsep}{1.5mm}
\centering
\scriptsize
\begin{tabular}{ccccccccccccc}
\hline\hline
CNAME &\ion{Sc}{I} &\ion{Sc}{II} &\ion{V}{I} &\ion{Cr}{I} &\ion{Cr}{II} & \ion{Mn}{I} & \ion{Fe}{I} &\ion{Fe}{II}
&\ion{Co}{I} &\ion{Ni}{I} &\ion{Cu}{I} & \ion{Zn}{I} \\
\hline
13364831-6206517&3.26$\pm$0.07 &3.43$\pm$0.09 &4.11$\pm$0.07&5.69$\pm$0.10 &5.71$\pm$0.10 &5.50$\pm$0.05 &7.53$\pm$0.07 &7.63$\pm$0.09 &4.98$\pm$0.08 &6.31$\pm$0.10 &4.11$\pm$0.06 &4.66$\pm$0.03\\	
13365597-6205130&3.19$\pm$0.06 &3.39$\pm$0.12 &4.06$\pm$0.09&5.64$\pm$0.11 &5.69$\pm$0.11 &5.48$\pm$0.05 &7.50$\pm$0.07 &7.59$\pm$0.07 &4.96$\pm$0.05 &6.30$\pm$0.11 &4.16$\pm$0.09 &4.51$\pm$0.01\\	
13365882-6205197&3.28$\pm$0.07 &3.37$\pm$0.23 &4.13$\pm$0.09&5.68$\pm$0.12 &5.69$\pm$0.16 &5.46$\pm$0.08 &7.45$\pm$0.09 &7.55$\pm$0.11 &4.94$\pm$0.12 &6.23$\pm$0.13 &4.13$\pm$0.10 &4.53$\pm$0.19\\	
13370214-6206095&3.22$\pm$0.07 &3.44$\pm$0.19 &4.09$\pm$0.08&5.69$\pm$0.10 &5.77$\pm$0.10 &5.48$\pm$0.09 &7.52$\pm$0.08 &7.66$\pm$0.08 &4.99$\pm$0.07 &6.32$\pm$0.09 &4.25$\pm$0.10 &4.55$\pm$0.01\\	
13370523-6206433&3.16$\pm$0.09 &3.32$\pm$0.09 &4.01$\pm$0.09&5.61$\pm$0.10 &5.73$\pm$0.05 &5.41$\pm$0.09 &7.44$\pm$0.08 &7.58$\pm$0.11 &4.91$\pm$0.06 &6.23$\pm$0.12 &4.20$\pm$0.14 &4.41$\pm$0.11\\	
13371182-6206030&3.31$\pm$0.06 &3.42$\pm$0.14 &4.18$\pm$0.08&5.69$\pm$0.11 &5.72$\pm$0.06 &5.53$\pm$0.06 &7.53$\pm$0.08 &7.58$\pm$0.09 &5.03$\pm$0.06 &6.33$\pm$0.10 &4.25$\pm$0.16 &4.44$\pm$0.10\\	
13365001-6205376&3.26$\pm$0.06 &3.42$\pm$0.13 &4.12$\pm$0.08&5.67$\pm$0.11 &5.71$\pm$0.09 &5.47$\pm$0.07 &7.53$\pm$0.07 &7.63$\pm$0.09 &4.97$\pm$0.05 &6.33$\pm$0.10 &4.19$\pm$0.09 &4.72$\pm$0.08\\	
\hline
\end{tabular}
\end{table*}

\begin{table*}
\caption{UVES members neutron-capture element abundances}
\label{t_7}
\setlength{\tabcolsep}{1.5mm}
\centering
\normalsize
\begin{tabular}{ccccccccc}
\hline\hline
CNAME      &\ion{Y}{II}  &\ion{Zr}{I} & \ion{Zr}{II}  &\ion{Ba}{II} &\ion{La}{II} &\ion{Ce}{II} &\ion{Nd}{II} &\ion{Eu}{II} \\
\hline
13364831-6206517&2.32$\pm$0.07&2.63$\pm$0.04 &2.84$\pm$0.10  &2.41$\pm$0.04 &1.12$\pm$0.08 &1.97$\pm$0.05  &1.71$\pm$0.08 &0.53$\pm$0.02 \\
13365001-6205376&2.32$\pm$0.08&2.63$\pm$0.05 &2.75$\pm$0.07  &2.40$\pm$0.04 &1.01$\pm$0.03 &1.86$\pm$0.01  &1.64$\pm$0.13 &0.55$\pm$0.02 \\
13365597-6205130&2.25$\pm$0.07&2.55$\pm$0.03 &2.69$\pm$0.05  &2.26$\pm$0.08 &1.09$\pm$0.07 &1.76$\pm$0.02  &1.65$\pm$0.08 &0.58$\pm$0.02 \\
13365882-6205197&2.20$\pm$0.14&2.66$\pm$0.05 &2.75$\pm$0.14  &2.27$\pm$0.08 &1.06$\pm$0.07 &1.65$\pm$0.05  &1.65$\pm$0.08 &0.53$\pm$0.02 \\
13370214-6206095&2.37$\pm$0.09&2.57$\pm$0.06 &2.80$\pm$0.08  &2.43$\pm$0.02 &1.09$\pm$0.05 &1.91$\pm$0.01  &1.70$\pm$0.10 &0.57$\pm$0.03 \\
13370523-6206433&2.22$\pm$0.08&2.50$\pm$0.06 &2.78$\pm$0.09  &2.31$\pm$0.03 &1.03$\pm$0.09 &1.72$\pm$0.02  &1.62$\pm$0.09 &0.51$\pm$0.02 \\
13371182-6206030&2.31$\pm$0.07&2.68$\pm$0.06 &2.79$\pm$0.08  &2.31$\pm$0.08 &1.17$\pm$0.08 &1.85$\pm$0.02  &1.73$\pm$0.07 &0.58$\pm$0.01 \\
\hline
\end{tabular}
\end{table*}

The redetermined parameter values given in the last column of  Table~\ref{t_1}  are in good agreement with previous studies. The reddening we found is compatible within the uncertainties with the values given by \citet{Piatti98} and \citet{Kharchenko13} and a bit lower than that derived by \citet{Tadross08}, who however does not give the associated uncertainty. Also the distance is in very good agreement with that given in \citet{Piatti98} and \citet{Kharchenko13}, while that found by \citet{Tadross08} is much smaller than the others. The ages given in the literature are a bit larger than that we found in the present work. We note that the larger age (still consistent within the errors) adopted by \citet{Piatti98} is due to their averaging the results obtained from isochrone fitting and integrated spectroscopy. Their photometry-based age is about 0.9 Gyr, in reasonable agreement with our findings. 

We also show a comparison between our best fit isochrone and the data in the ($\log T_\mathrm{eff}$, $\log g$) plane (see Fig. 6). The relatively large scatter displayed by the MS stars is expected as in this case the atmospheric parameters (given in Table~\ref{tab_atmpar}) were derived from GIRAFFE low resolution spectra. Despite this scatter,  the best isochrone is fully compatible with the data for both MS and RC stars (for which the atmospheric parameters are much better constrained, as they are based on high resolution UVES spectra).

The location of the RC stars on the CMDs strongly affects the distance determination and limits the acceptable values of $DM_0$. The uncertainty in $DM_0$ is asymmetric around the best value. The best value of the distance modulus is obtained when using the upper part of the central helium burning phase.  Adopting a larger distance would cause these RC stars to move away from the isochrone, thus producing a worse fit. Another possible solution would be to fit such stars with the lower part of the theoretical RC, which is still acceptable in the fitting procedure and gives a relatively good quality fit. This latter case corresponds to a smaller distance. Thus, this simple statement should qualitatively explain the asymmetry in the uncertainty on $DM_0$. 
Another point concerns the large discrepancy between the parallax provided by Gaia DR2 and what we derived here (similar to that used in the literature). We found a mean parallax of $0.406^{+0.047}_{-0.019}$~mas (corresponding to  2.47~pc), to be compared with the $0.338\pm0.043$~mas (corresponding to 2.94~kpc) we derived using the Gaia parallaxes. The difference between the two determinations is about 0.068~mas. We tried to use the Gaia distance as a prior in our isochrone fitting, but the results were not satisfactory. The isochrones tended to drastically underestimate the magnitude of the MS stars, while the Turn-Off region and the RC were not well reproduced. Therefore, it seems very unlikely that the cluster could have such a large distance. It is noted that similar discrepancies have been reported for distant clusters
 \citep{lindegren2018}. Such systematic effects are expected to be minimized at the end of the Gaia mission.

\begin{table}
\caption{Reference element abundances.}
\label{t_8}
\setlength{\tabcolsep}{0.05in}
\centering
\begin{tabular}{l r  r r r }
\hline\hline
Element & Atomic & Grevesse(07)&i{\sc dr5}-solar&M67\\
        & number &       &      &  \\
\hline

\ion{Na}&	11	&6.17$\pm$0.04 &6.17$\pm$0.05 &\\
\ion{Mg}&	12	&7.53$\pm$0.09 &7.51$\pm$0.07 &\\
\ion{Al}&	13	&6.37$\pm$0.06 &6.34$\pm$0.04 &\\
\ion{Si}&	14	&7.51$\pm$0.04 &7.48$\pm$0.06&\\
\ion{Ca}&	20	&6.31$\pm$0.04 &6.31$\pm$0.12&\\
\ion{Sc}&	21	&3.17$\pm$0.10 &3.27$\pm$0.06 &\\
\ion{Ti}&	22	&4.90$\pm$0.06 &4.90$\pm$0.08 &\\
\ion{V}&	23	&4.00$\pm$0.02 &4.00$\pm$0.09 &\\
\ion{Cr}&	24	&5.64$\pm$0.10 &5.61$\pm$0.09 &\\
\ion{Mn}&	25	&5.39$\pm$0.03 &5.39$\pm$0.06 &\\
\ion{Fe}&	26	&7.45$\pm$0.05 &7.47$\pm$0.06 &\\
\ion{Co}&	27	&4.92$\pm$0.08 &4.83$\pm$0.08 &\\
\ion{Ni}&	28	&6.23$\pm$0.04 &6.23$\pm$0.07 &\\
\ion{Cu}&	29	&4.21$\pm$0.04 &4.12$\pm$0.10 &\\
\ion{Zn}&	30	&4.60$\pm$0.03 &4.60$\pm$0.06 &\\
\ion{Y}&	39	&2.21$\pm$0.02 &2.19$\pm$0.12 &\\
\ion{Zr}&	40	&2.58$\pm$0.02 &2.53$\pm$0.13 &\\
\ion{Ba}&	56	&2.17$\pm$0.07 &2.17$\pm$0.06 &\\
\ion{La}&	57	&1.13$\pm$0.05 &              &0.97$\pm$0.07\\
\ion{Ce}&	58	&1.70$\pm$0.10 &1.70$\pm$0.11 &\\
\ion{Nd}&	60	&1.45$\pm$0.05 &              &1.56$\pm$0.02\\
\ion{Eu}&	63	&0.52$\pm$0.06 &              &0.42$\pm$0.01 \\
\hline
\end{tabular}
\end{table}

\section{The chemical composition of Pismis 18} \label{abuns}

Tables~\ref{t_5}, \ref{t_6}, \ref{t_7} give elemental abundances in the form 12+log(X/H) of six high
confidence radial velocity members (plus the suspect binary) observed with UVES for light, $\alpha$,
Fe-peak, and neutron-capture elements. For each star, its elemental abundances
were computed by combining the results of different nodes for each absorption
line. The results of all lines of the same element are then combined to produce
the final abundance per element per star.  The corresponding errors were computed from
line-by-line abundance variations, after combining the abundances of the various
nodes (see \citealt{smiljanic14} for a detailed description).

In Table 7 we show the Solar reference abundances by \citet{grevesse07}, the GES i{\sc dr5} Solar
abundances and the median abundances of M67 giant stars. To compute the [X/H] ratios we adopted the Solar abundance scale of GES i{\sc dr5}, using for most elements the
homogenized Solar abundances provided in the final i{\sc dr5} table.  For elements that
are not measured in the Solar spectra (Nd and Eu) and for La, we adopted the abundances of member giant stars
in the calibration open cluster M67, which is known to have a chemical
composition very close to Solar \citep[see, e.g.][]{Tautvaisiene2000, Shetrone2000, Yong05, Randich06, Pace08,  Onehag14, Bertelli-Mota17}. 

The median abundances of Pismis~18 (for 22 elements, i.e. other than Li)  and their standard deviations based on the six member stars are provided in Col. 2 of Table 8. In Cols. 3 and 4, we present the median
[X/H] and [X/Fe] abundance ratios of Pismis~18 (the latter were calculated using \ion{Fe}{I} and not [Fe/H], see discussion in Sect.~5).

\begin{table}
\caption{Cluster average element abundances.}
\label{t_9}
\setlength{\tabcolsep}{0.05in}
\centering
\begin{tabular}{l r r r}
\hline\hline
Element  &Abundance& [X/H]&[X/Fe] \\
\hline
\ion{Na}{I}	& 6.50$\pm$0.02	& 0.33$\pm$0.05	&0.31$\pm$0.11\\
\ion{Mg}{I}	& 7.59$\pm$0.02	& 0.08$\pm$0.07 &0.04$\pm$0.16\\
\ion{Al}{I}	& 6.44$\pm$0.01	& 0.10$\pm$0.04	&0.08$\pm$0.13\\
\ion{Si}{I}	& 7.68$\pm$0.01	& 0.20$\pm$0.06	&0.16$\pm$0.13\\
\ion{Ca}{I}	&6.45$\pm$0.01	& 0.14$\pm$0.12 &0.11$\pm$0.13\\
\ion{Sc}{II}&3.41$\pm$0.05	& 0.14$\pm$0.08	&0.10$\pm$0.16\\
\ion{Ti}{I}	&4.91$\pm$0.01	& 0.00$\pm$0.08	&-0.05$\pm$0.13\\
\ion{V}{I}	&4.10$\pm$0.01	& 0.10$\pm$0.09	&0.04$\pm$0.13\\
\ion{Cr}{I}	&5.69$\pm$0.01	& 0.08$\pm$0.09 &0.03$\pm$0.15\\
\ion{Mn}{I}	&5.48$\pm$0.02   & 0.09$\pm$0.06	&0.06$\pm$0.12\\
\ion{Fe}{I}	&7.51$\pm$0.01	& 0.04$\pm$0.06	&\\
\ion{Fe}{II}&7.59$\pm$0.01	& 0.08$\pm$0.07	&\\
\ion{Co}{I}	&4.97$\pm$0.02	& 0.14$\pm$0.08	&0.11$\pm$0.12\\
\ion{Ni}{I}	&6.31$\pm$0.01	& 0.07$\pm$0.07	&0.03$\pm$0.15\\
\ion{Cu}{I}	&4.18$\pm$0.03	& 0.06$\pm$0.11	&0.05$\pm$0.14\\
\ion{Zn}{I}	&4.52$\pm$0.07	& 0.06$\pm$0.07	&-0.11$\pm$0.12\\
\ion{Y}{II}	&2.38$\pm$0.02	& 0.09$\pm$0.12	&0.06$\pm$0.13\\  
\ion{Zr}{II}&2.79$\pm$0.03	& 0.26$\pm$0.13	&0.23$\pm$0.13\\
\ion{Ba}{II}&2.31$\pm$0.03	& 0.14$\pm$0.07 &0.15$\pm$0.12\\
\ion{La}{II}&1.09$\pm$0.01	& 0.12$\pm$0.06	&0.09$\pm$0.13\\
\ion{Ce}{II}&1.80$\pm$0.02	& 0.10$\pm$0.11	&0.07$\pm$0.10\\
\ion{Nd}{II}&1.68$\pm$0.01	& 0.12$\pm$0.01	&0.09$\pm$0.13\\
\ion{Eu}{II}&0.55$\pm$0.01	& 0.13$\pm$0.01	&0.11$\pm$0.10\\
\hline
\end{tabular}
\end{table}

\subsection{Light elements}
\subsubsection{Lithium}
Lithium was measured in stars along the MS and in all RC stars. Figure~\ref{fig7} gives the dependence of Li abundance on \teff, that is on evolutionary phase. All Li-rich stars, and all upper limits, are MS stars. Instead,
all of the six evolved members have true measurements, not upper limits. Their values are all low, as expected from their evolutionary phase, after the first dredge-up. The star just below the RC and the probable binary have the highest Li abundances
(Table~\ref{t_5}) even if they do not qualify as 'Li-rich' in the  absolute sense \citep[see e.g. the Li-rich stars discussed by][discovered among GES targets]{Smiljanic2018b}

\subsubsection{Sodium}
Sodium is the only elemental abundance which is significantly super-solar,  with [Na/Fe]=+0.31. However, this is the value in LTE, and NLTE corrections for RC stars are about -0.1 dex \citep[see, e.g.][]{Smiljanic16}.  A similar overabundance in giant stars (it is noted that the UVES targets are evolved stars) has been routinely observed and can be attributed to evolutionary mixing of products of the Ne-Na cycle during the first dredge-up phase. \citet{Smiljanic16} discuss this using GES clusters, showing that a value of about 0.2 dex is normal for clusters of similar age/turnoff mass (see their Table 2 and Fig.~5).

\begin{figure}
\centering
\includegraphics[scale=0.33]{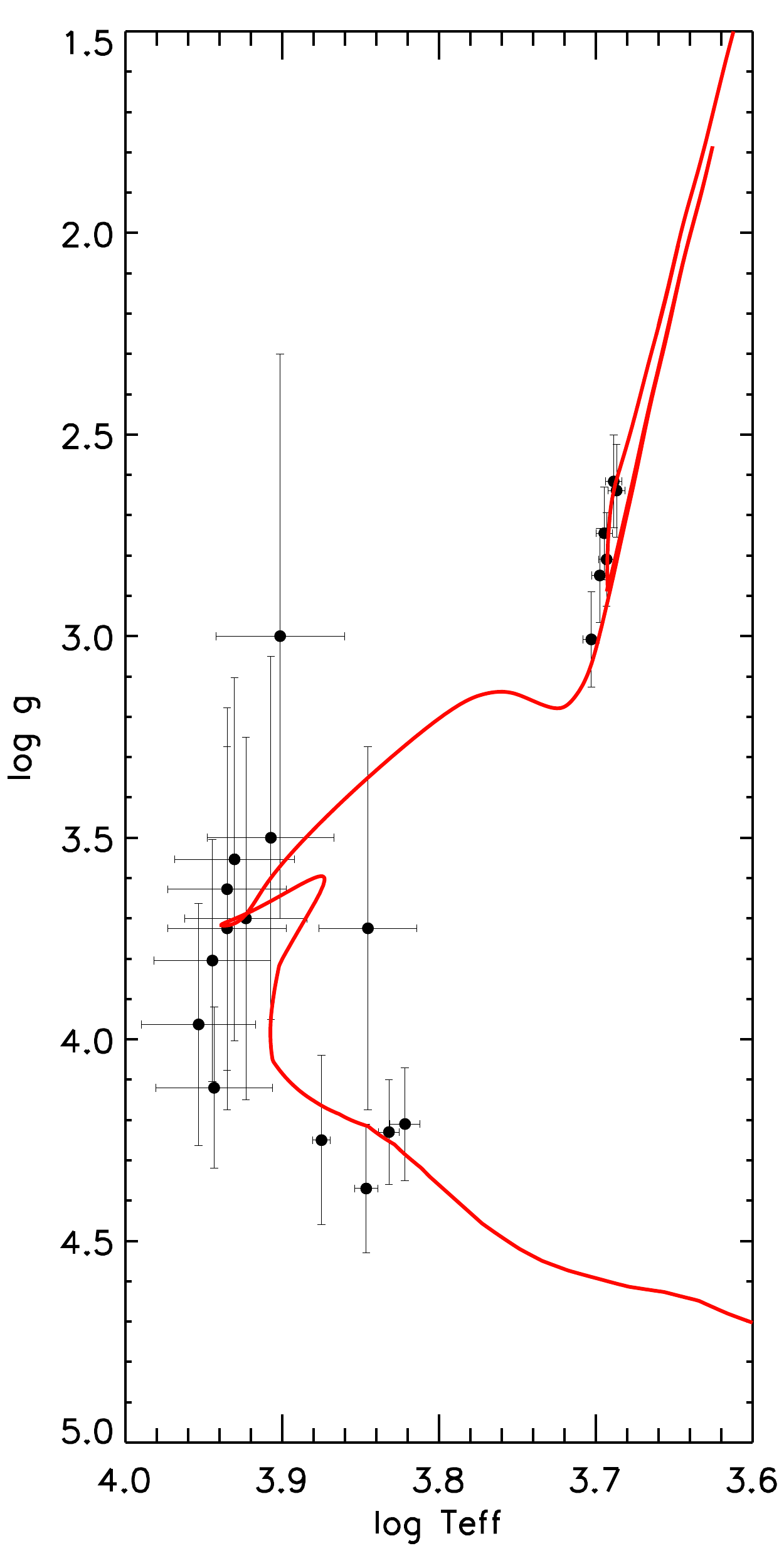}
\caption{Comparison between the Pisa best fit isochrone (corresponding to $\tau = 700^{+40}_{-50}$~Myr, $E(B-V)=0.562^{+0.012}_{-0.026}$~mag and $DM_0=11.96^{+0.10}_{-0.24}$~mag) and the determined $\log T_\mathrm{eff}$ and $\log g$ values.}
\label{fig:logte_logg}
\end{figure}
\begin{figure}
\centering
\includegraphics[width=0.7\hsize]{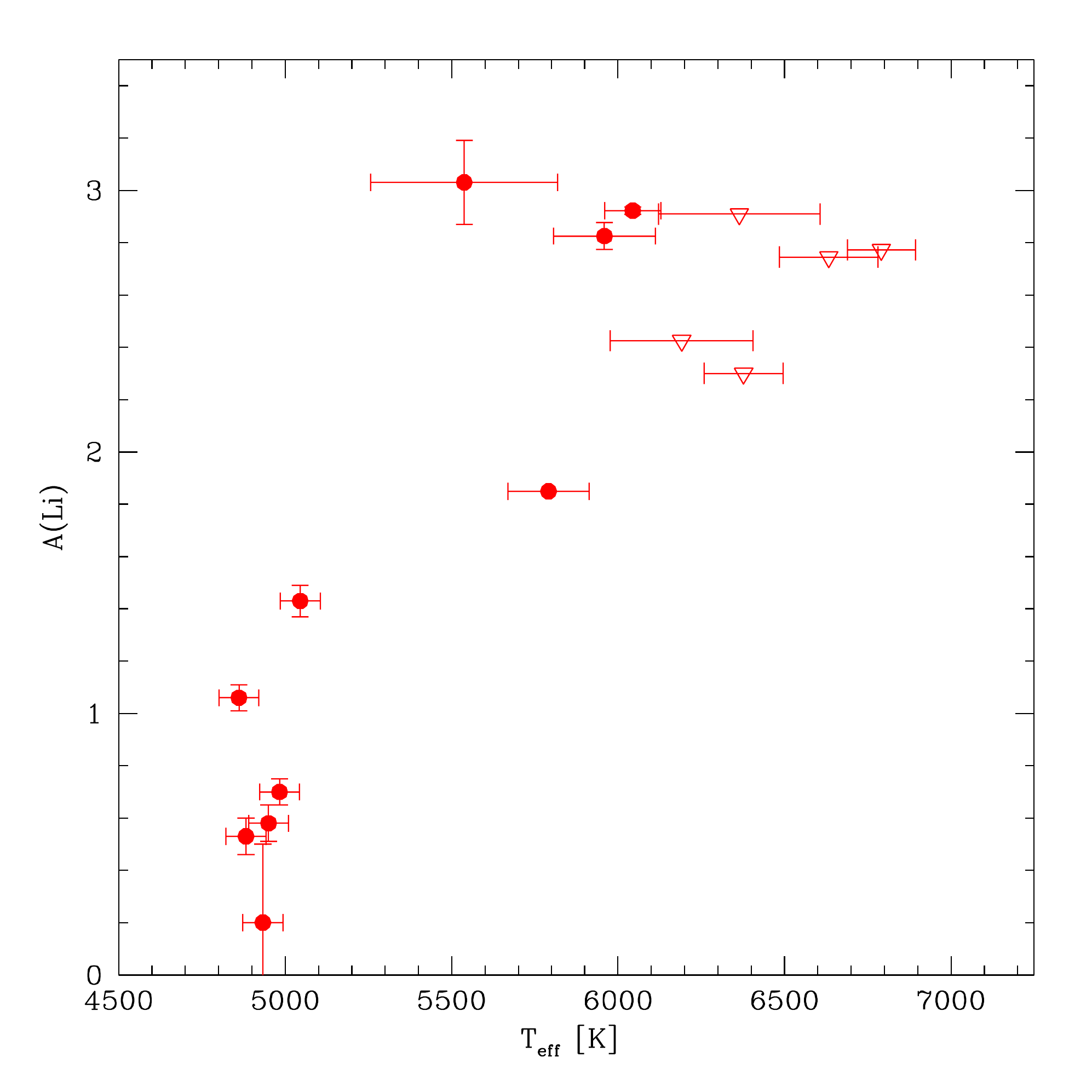}
\caption{Abundance of Li as a function of temperature (filled symbols with error bars indicate specific measured values, while empty triangles indicate upper limits.}
\label{fig7}
\end{figure}

\subsection{$\alpha$-elements}
As far as $\alpha$-elements (Mg, Si, Ca, Ti - even if only the first traces exclusively massive-stars nucleosynthesis, while others, especially Ti, come from several channels) are concerned, 
the abundances over H measured for Pismis~18 are
close to solar values or slightly over-abundant (see Table~8, for instance
[\ion{Si}{I}/H]=0.20$\pm$0.06). 
Also [Mg/Fe], [Ca/Fe], [SI/Fe] and [Ti/Fe] range from slightly sub-solar to super-solar, with a mean value of [$\alpha$/Fe]=0.07$\pm$0.13. The individual 
abundances and abundance ratios are provided in Table~8.

Due to the different
timescales involved in the production of the $\alpha$ elements and of iron and
to the different star formation rates in the inner and outer part of the
Galactic disc –the so-called inside-out formation of the disc-  classical
chemical evolution models predict a depletion of the [$\alpha$/Fe] ratio in the
inner disc coupled with an enhancement in the outskirts.  
This is indeed
predicted by several chemical evolution models \citep[see, e.g.][]{Magrini09,Magrini15,Kubryk13,Minchev14}. For this reason, it would be expected that clusters located in the inner disc should present a depletion in
$\alpha$-elements over iron with respect to solar values. However, observations
of young populations, as shown for instance  by \citet{Chiappini2015}, \citet{Martig2015}, \citet{Magrini17} and \citet{Casamiquela2018}  seem to
contradict the expectations of chemical evolution models built on an inside-out
scenario \citep[see for example Fig.~10 of][]{Minchev14}. 
However, the point is that these results do not invalidate the inside-out formation of the disc, but indicate that the nucleosynthesis of some elements is more complex than believed in the past. 
In particular, \citet{chiappini05} already pointed out the Mg problem 
concluding that larger quantities of Mg should have been produced at recent epochs to explain the trend of [Mg/Fe] vs. [Fe/H]. 
This can be achieved either by different SNIa models or by SNII metallicity dependent yields, as done in \citet{Magrini17}. 
Pismis~18, which is located at $\simeq$6.8 kpc from the Galactic Centre,
allows us to check the presence or not of such depletion. With [$\alpha$/Fe]=0.07$\pm$0.13 it is in agreement,
within the errors,  with both the model predictions and the observations of the
GES {\sc dr4} clusters shown in Fig.~8 of \citet{Magrini17}.

\subsection{Iron-peak elements}
\begin{figure}[t]
\centering
\includegraphics[width=0.9\hsize]{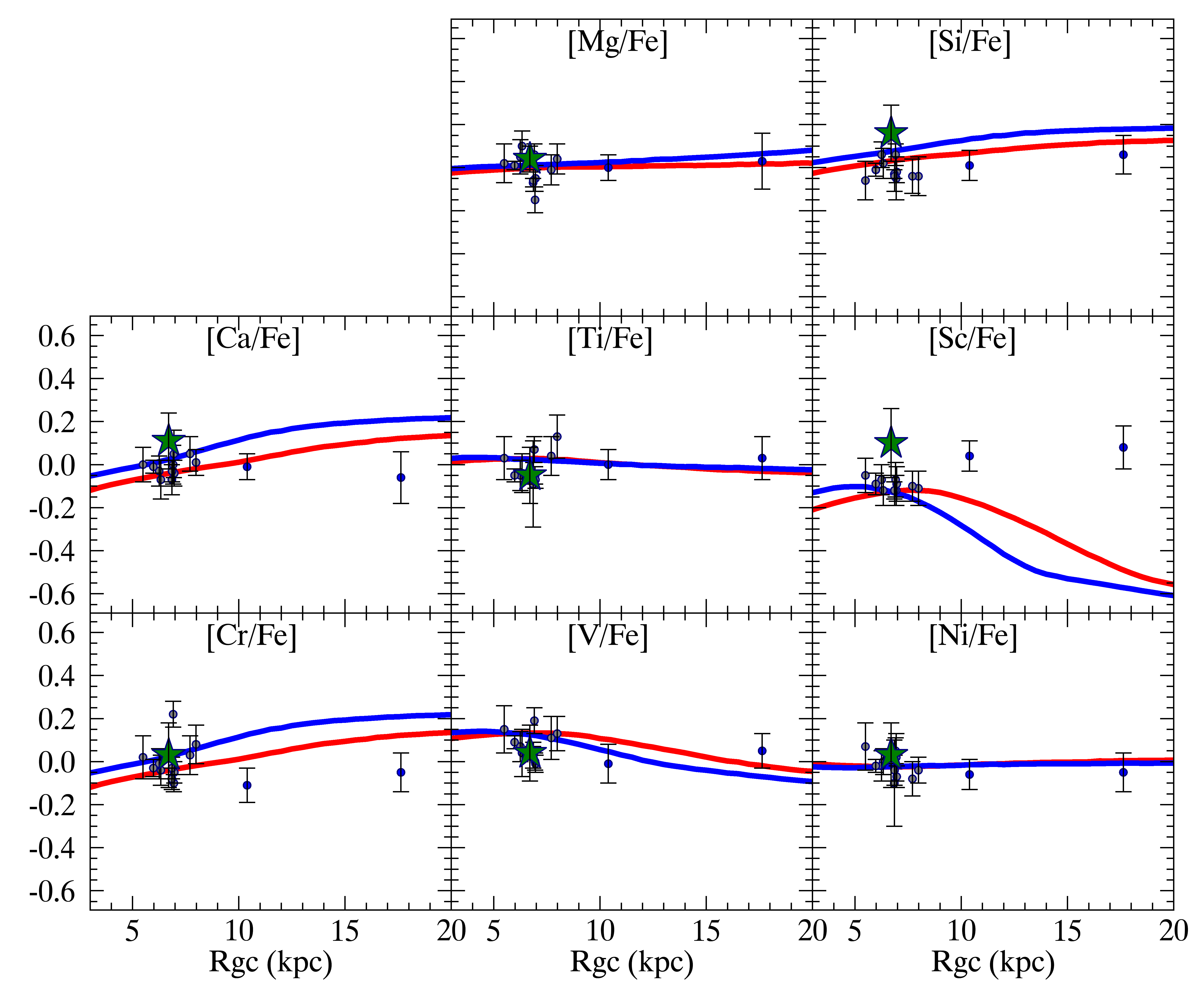}
\caption{Abundance ratios as a function of galactocentric distance ($R_{GC}$) for Pismis~18 (green star) compared with the results of \citet{Magrini17} for both clusters (grey circles --the youngest clusters, age$<$2 Gyr, and in blue the oldest ones) 
and models (red--present time-- and blue -- 5~Gyr ago -- curves).}
\label{fig_m17}
\end{figure}

We measured the abundances of several
iron-peak elements such as Sc, V, Cr, Mn, Co, and Ni, whose abundance ratios are
all roughly solar. 
For some of them we could compare the theoretical and observational
results of \citet{Magrini17}. 
In Fig.~\ref{fig_m17}, we compare the abundance ratios of Pismis~18 with those presented in \citet{Magrini17} and with the results of their chemical evolution model.
The agreement is good for Cr, V and Ni, while the model predictions for [Sc/Fe] are not able to explain the
observational results.

\begin{figure*}
\centering
\sidecaption
\includegraphics[width=0.7\hsize]{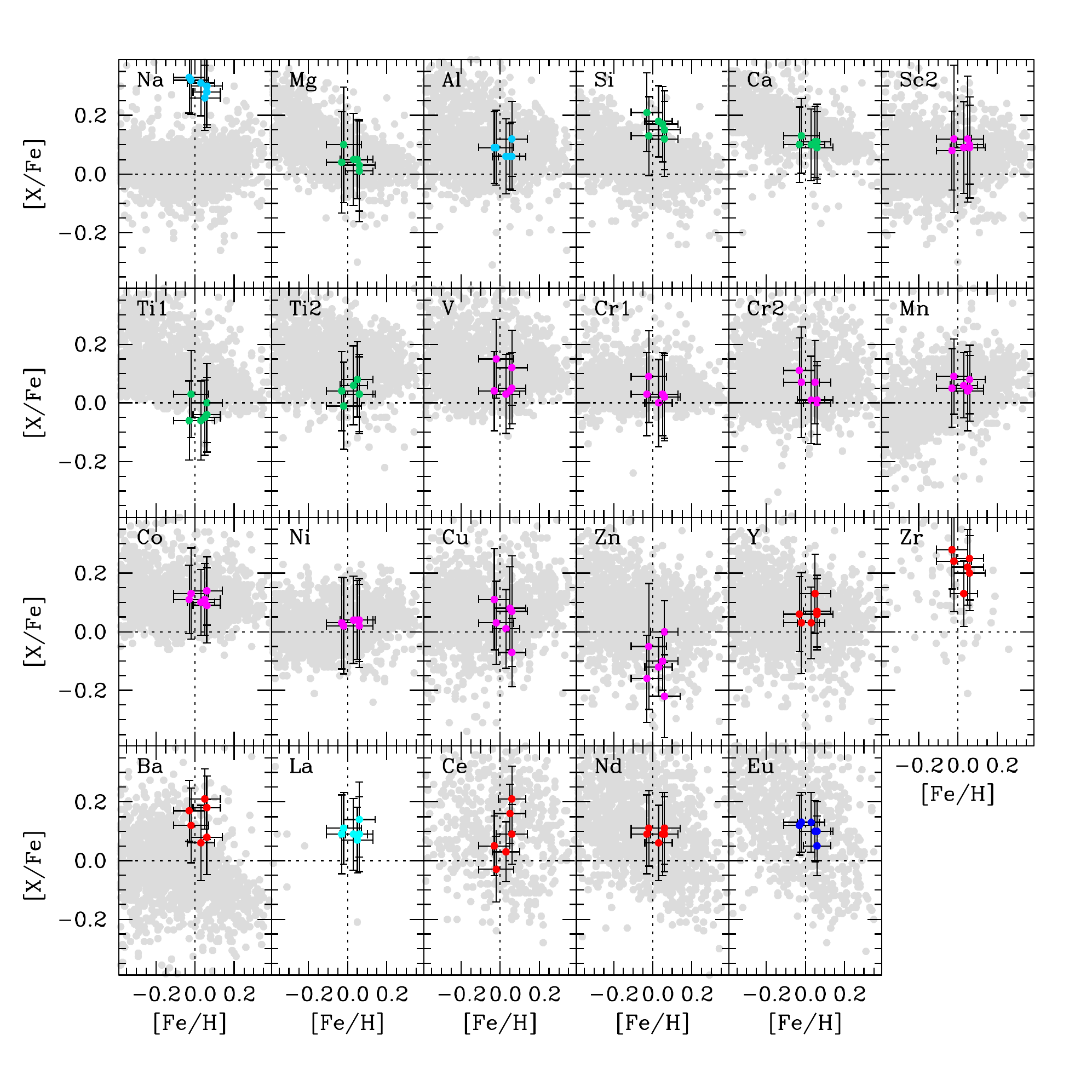}
\caption{ Distribution of [X/Fe] ratios as a function of [Fe/H] for 23 species including light (cyan), $\alpha$ (green), Fe-peak (magenta) and n-capture (red for s-process and blue for r-process) elements. The grey points correspond to field MW stars in GES (see text; it is noted that no Y measurements are available for them).}
\label{fig9}
\end{figure*}

\subsection{Heavy elements}
The slow-(s-) and rapid-(r-) neutron-capture elements are slightly
enhanced, with [\ion{Y}{II}/Fe]=0.06$\pm$0.10, [\ion{Zr}{II}/Fe]=0.20$\pm$0.07,
[\ion{Ba}{II}/Fe]=0.08$\pm$0.12, [\ion{La}{II}/Fe]=0.09$\pm$0.13, and [\ion{Ce}{II}/Fe]=0.07$\pm$0.10 for the
s-process elements and [\ion{Nd}{II}/Fe]=0.09$\pm$0.13 and [\ion{Eu}{II}/Fe]=0.11$\pm$0.10 for the
r-process elements. 
The abundances of neutron capture elements for Pismis~18 are also included
in the \citet{Magrini18}  (hereafter M18) study. The small differences between 
the results presented here and in M18 are related to the more strict selection 
of stars included to compute the average value by M18, 
in which stars with large errors on individual abundances 
were not considered. In addition, for Eu, M18 adopted 
the solar value from Grevesse et al. (2007).
An increase of the abundance of slow neutron capture
elements is expected in the youngest stellar populations \citep{dorazi09,
maiorca12, spina17} due to the strong contribution of low mass Asymptotic Giant
Branch (AGB) stars, which, given their long lifetimes, restore their material to the interstellar medium at late times and hence can be incorporated only in the youngest generation of new born stars.
In M18, we have studied the effect of the use Fe~II instead of Fe~I to compute the abundances for elements with singly ionized atoms, which, in principle, 
would be more appropriate. 
However, there is a remarkable agreement between log(Fe~I/H) and log(Fe~II/H) in 
GES i{\sc dr5} samples. Thus, since the Fe~II abundances are affected by larger 
uncertainties, we adopted in this paper the Fe~I abundances to compute the [X/Fe] abundance ratios.

\begin{figure}
\centering
\includegraphics[width=0.9\hsize]{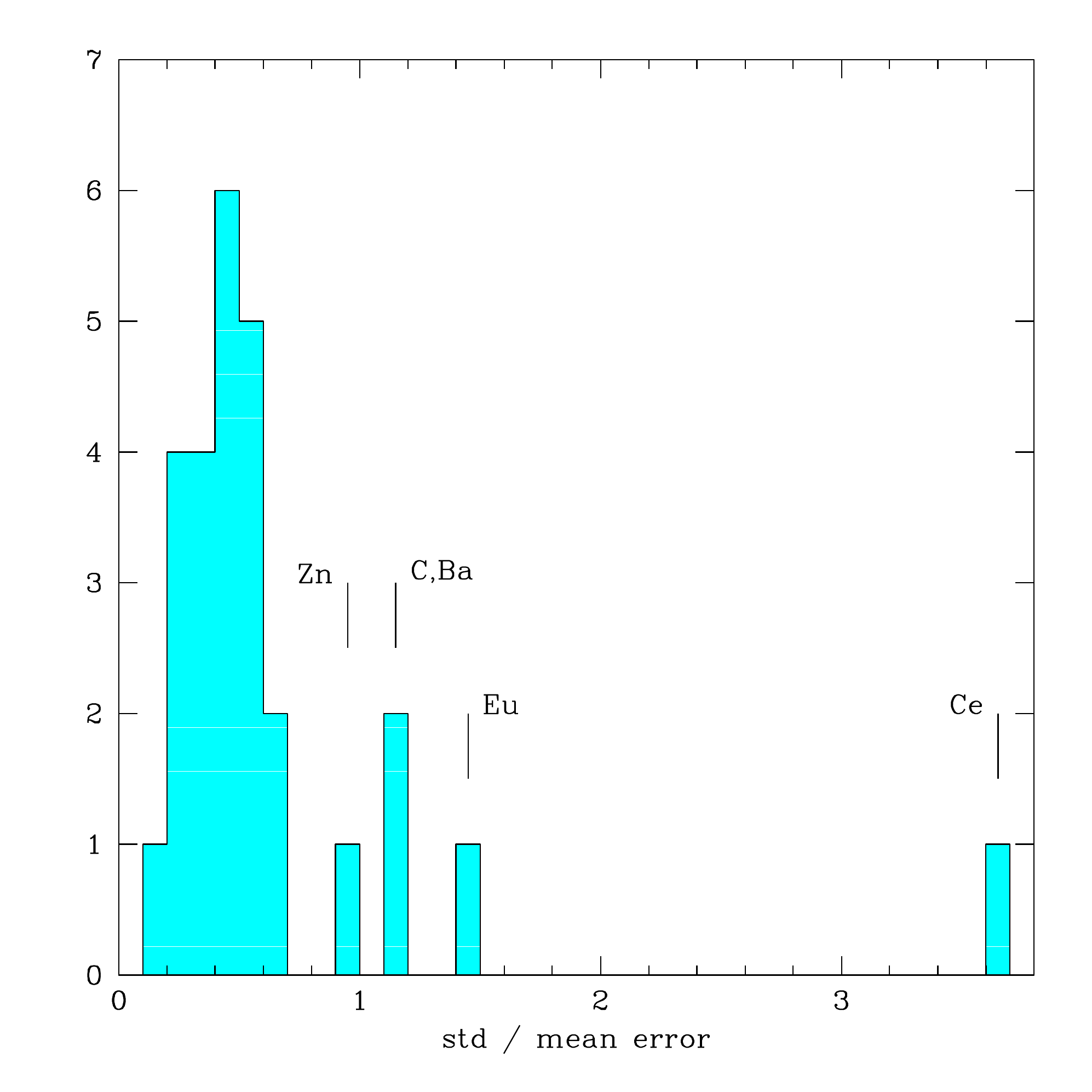}
\caption{Histogram of the ratio between average error and standard dispersion for all species (Li excluded) measured for the high probability members of Pismis~18.}
\label{fig10}
\end{figure}

As stated in the Introduction, OCs are considered to be simple stellar populations. This implies that all stars in the cluster share the same initial chemical composition and that any differences found should be attributed solely to evolutionary effects (e.g. diffusion, mixing) or to the presence of a companion. This homogeneity in chemical composition prompted the suggestion of chemical tagging, that is, of identifying the common origin of apparently unrelated  stars (see e.g. the review
by \citealt{freeman02}). While no dedicated study, such as the one by \citet{ness18} on several APOGEE open clusters, has been performed for the GES cluster sample yet, we have not found any compelling indication of unexplained intrinsic spreads \citep[see e.g.][just to cite M11, Trumpler~20, and Trumpler~23]{CantatGaudin14,Donati14,Overbeek17}. The case of Pismis~18, although based only on six stars, does not differ from this trend, as it shows significant homogeneity in all elemental abundances: the error on the mean abundance (see Col. 2 of Table 9) ranges from 0.01~dex for several elements (e.g. \ion{Al}{I}, \ion{Si}{I}, \ion{Ca}{I}, \ion{Ti}{I}, \ion{Fe}{I}, \ion{Fe}{II}),
to 0.02-0.04 dex (e.g. for \ion{Ti}{II}, \ion{Cr}{II}, \ion{Ba}{II}), and rarely is larger than 0.05 dex. The dispersion slightly increases when introducing the conversion to elemental ratios with respect to H and Fe (Cols. 3 and 4 of Table 9).  As shown in Fig.~\ref{fig9}, none of the 23 species shows any significant dispersion (see the error bars, indicating the average error on each star).
As a further check, we plot in Fig.~\ref{fig10} the ratio of the rms of the log[X/H] values and the average error, using the values in Tables 4, 5 and 6  for the calculations. In almost all cases the dispersion is smaller than the error, indicating homogeneity in composition. A few discrepant cases are indicated,  with ratios larger than 1. Generally, this occurs for elements difficult to measure (e.g. Ba, with very strong lines, or Ce, with only a few weak lines).

Figure~\ref{fig9} also shows  the distribution of elemental ratios as a function of iron abundance for field MW stars observed by GES, for homogeneous comparison. We selected only stars observed with UVES (setup U580) and in the metallicity range -0.4 to 0.4; there are about 1700 stars, almost all dwarfs (the giants are about 20 in total). This paucity of giants explains the large difference ($>0.2$ dex) we see for [Na/Fe]; in fact, giants of mass about 2-2.5 M$_\odot$, such as those in Pismis~18, are expected to show enhanced Na abundance with respect to MS stars, due to mixing (see e.g. \citealt{Smiljanic16,smiljanic2018a} for a discussion). For all other cases the distributions 
of cluster values fall within the range of field values.

\section{Summary and conclusions}
We have conducted an extended radial velocity and proper motion membership study as well as spectroscopic metallicity measurements for 142 stars in and around the inner disc Galactic cluster Pismis~18, using Gaia-ESO Survey i{\sc dr5} data, as well as Gaia DR2 data. Of the 142 stars, we could confirm high confidence membership for 26 stars, out of which six lie on the red clump and 20 on the upper MS of the cluster.  These stars were used to determine the systemic velocity of the cluster,  $-27.5\pm2.5$ \kms. Gaia DR2 photometry was used to re-determine cluster parameters based on high confidence member stars only. According to these new estimates, Pismis~18 has an age of $\tau = 700^{+40}_{-50}$~Myr, interstellar reddening of $E(B-V)=0.562^{+0.012}_{-0.026}$~mag and a de-reddened distance modulus $DM_0=11.96^{+0.10}_{-0.24}$~mas (corresponding to $2.47^{+0.11}_{-0.26}$~kpc). Using abundance measurements for 22 (i.e. not including Li) elements based on  high-resolution spectra of the six radial-velocity member stars on the red clump, we determined that the  median metal abundance of Pismis~18 is above solar (using all measured elements) at $0.23\pm0.05$~dex, with the ratio of the $\alpha$ elements to iron about solar (within the errors) at [$\alpha$/Fe]$=0.07\pm0.13$.  A slight enhancement was observed for neutron-capture elements, which is expected for younger disc populations.

\begin{acknowledgements}
Based on data products from observations made with ESO Telescopes at the La
Silla Paranal Observatory under programme ID 188.B-3002. These data products
have been processed by the Cambridge Astronomy Survey Unit (CASU) at the
Institute of Astronomy, University of Cambridge, and by the FLAMES/UVES
reduction team at INAF/Osservatorio Astrofisico di Arcetri. These data have been
obtained from the Gaia-ESO Survey Data Archive, prepared and hosted by the Wide
Field Astronomy Unit, Institute for Astronomy, University of Edinburgh, which is
funded by the UK Science and Technology Facilities Council.

This work was partly supported by the European Union FP7 programme through ERC
grant number 320360 and by the Leverhulme Trust through grant RPG-2012-541. We
acknowledge the support from INAF and Ministero dell' Istruzione, dell'
Universit\`a' e della Ricerca (MIUR) in the form of the grant 'Premiale VLT
2012'. The results presented here benefit from discussions held during the
Gaia-ESO workshops and conferences supported by the ESF (European Science
Foundation) through the GREAT Research Network Programme.

The research leading to these results has received funding from the European
Community's Seventh Framework Programme (FP7-SPACE-2013-1) under grant agreement
no. 606740.

This work has made use of data from the European Space Agency (ESA) mission
{\it Gaia} (\url{https://www.cosmos.esa.int/gaia}), processed by the {\it Gaia}
Data Processing and Analysis Consortium (DPAC,
\url{https://www.cosmos.esa.int/web/gaia/dpac/consortium}). Funding for the DPAC
has been provided by national institutions, in particular the institutions
participating in the {\it Gaia} Multilateral Agreement.

E.~T. acknowledges the University of Pisa (\emph{Low and intermediate mass stellar
models for the age determination of stellar clusters observed by the Gaia
satellite}, PI: S.Degl'Innocenti) and INFN (\emph{Iniziativa specifica TAsP}).
A.~R.~C. acknowledges support through the Australian Research Council through grant DP160100637.
E.~D.~-M. acknowledges the support from Funda\c{c}\~ao para a Ci\^encia e a Tecnologia (FCT) through national funds
and from FEDER through COMPETE2020 by the following grants UID/FIS/04434/2013 \& POCI--01--0145-FEDER--007672,
PTDC/FIS-AST/7073/2014 \& POCI--01--0145-FEDER--016880 and by the Investigador FCT contract IF/00849/2015.
S.~F. was supported by the project grant “The New Milky Way” from Knut and Alice Wallenberg Foundation.
R.~S. acknowledges support from the Polish Ministry of Science and Higher Education. A.~B. acknowledges PREMIALE 2015 MITiC (PI B. Garilli).

\end{acknowledgements}

\bibliographystyle{aa} 
\bibliography{biblio_pismis18.bib}

\end{document}